\documentclass[12pt]{article}

\usepackage{graphicx}
\usepackage{a4wide}

\usepackage{mathrsfs}
\usepackage{amsmath}
\usepackage{amssymb}
\usepackage{amsfonts}

\def \dels {\partial\kern-.6em / \kern.1em}

\usepackage{dcolumn}

\begin{document}

\title{QCD landscape?}
\author{
Cong-Xin Qiu
~\footnote{E-mail: congxin.qiu@gmail.com; Homepage: http://oxo.lamost.org/} }
\date{~}

\maketitle

\begin{abstract}

Just comparing with the scenario that the $(3+1)$-dimensional ``real
world'' of the Calabi-Yau compactification has a tremendous
landscape, we conjecture that a $(4+1)$-dimensional holographic
theory may also hold a landscape of its vacua. Unlike the
traditional studies of the AdS/CFT phenomenology where the vacua are
always constructive, we discuss the proper holographic vacua and
their flux compactification, starting from some general compact
Einstein manifolds. The proper vacua should be restricted by (i) a
consistent worldsheet theory that possesses the superconformal
symmetry, (ii) some definite symmetries to keep/break the
corresponding symmetries of the dual field theory, (iii) certain
brane/flux configurations to cancel anomalies, and (iv) stabilities.
We consider diverse fundamental parameters of the dual field theory,
fixed by some special vacuum moduli.

In an opposite way, if some field theory such as QCD holds an AdS
dual, it may also possesses various fundamental parameters by an
``landscape'' of its vacuum. Different vacua may be adjacent with
each other, and divided by domain walls. If the size of a single
vacuum region is smaller than the visible universe, it may be
testable. We discuss the consequences of this conjecture in the
astrophysical environments, include but not limit to: (i)
consistency with the critical energy density of the universe, (ii)
the behaviors of cosmic rays, (iii) the stability and abundance of
deuterons and other nuclei in the big-bang nucleosynthesis and the
star burning scenarios, and (iv) the existence of strange/charm
stars.

\vspace{0.5cm}
\end{abstract}

\clearpage

\section{Introduction\label{sec:introduction}}

The anti-de Sitter/conformal field theory (AdS/CFT) correspondence,
one of the most ambitious scenarios in string phenomenology,
conjectures that a type IIB superstring theory on $\mathrm{AdS}_5
\times S^5$ is equivalent with a $\mathcal{N} = 4$ $U(N_c)$ super
Yang-Mills (SYM) theory in four-dimensions~\cite{Maldacena:1997re},
or more generally a gravity theory on $\mathrm{AdS}_{p+2} \times
\mathcal{M}_q$ is dual to a $(p+1)$-dimensional boundary
CFT~\cite{Witten:1998qj,Aharony:1999ti}. The idea of
``holography''~\cite{Susskind:1998dq} also pushes the applications
of AdS/CFT to more realistic environments, such as
QCD~\cite{Peeters:2007ab,Mateos:2007ay}, or condensed matter
systems~\cite{Herzog:2009xv}.

Nevertheless, the compact dimensions (thus the ``vacua'') and their
stabilization in AdS/CFT models, has seldom been studied
systematically. On the one hand, theoretical researches always study
some specific vacuum by constructive methods. For example, they
break boundary supersymmetry by quotient spaces $S^5 /
\Gamma$~\cite{Kachru:1998ys,Lawrence:1998ja}, or the conifold
construction~\cite{Klebanov:1998hh,Acharya:1998db,Morrison:1998cs}.
On the other hand, phenomenological models which aim to approach the
``real world'' physics, always neglect the discussions of the
compact dimensions directly. However, although difficult, the study
of AdS/CFT vacua has no alternative but within the framework of flux
compactification~\cite{Grana:2005jc,Douglas:2006es}. Some founding
works in this direction can be found
in~\cite{Aharony:2008wz,Polchinski:2009ch}.

The original studies of flux compactification, always aim to the
Calabi-Yau threefolds. The main reason is that $\mathrm{CY}_3$,
which possesses a special holonomy of $SU(3) \subset SO(6)$, can
reduce the ten-dimension critical superstring theory to some
four-dimensional effective field theory which possesses $\mathcal{N}
= 1$ supersymmetry. One of the properties of this scenario beyond
one's expectation, is the tremendously abundant vacua, which mainly
rise from the not-very-small Betti numbers $b_2$ and $b_3$ of
$\mathrm{CY}_3$, and the various possible fluxes wrapped on it; this
set of vacua is always called a ``string
landscape''~\cite{Susskind:2003kw}. Different vacua in the landscape
hold different fundamental parameters. It was argued that the number
of consistent quasirealistic flux vacua may be greater than
$10^{500}$~\cite{Douglas:2003um}, and models has been constructed to
solve the cosmological constant problem using this
property~\cite{Bousso:2000xa}.

In this paper, we conjecture that as an analog, a holographic theory
may also hold a landscape of vacua. We verify this hypothesis in two
different ways, from top-down and from bottom-up. Along the first
root, we discuss properties of the compact manifolds, and the
restrictions of them from physical purposes. For the uncompactified
dimensions to be AdS, the compact manifold should be Einstein; thus,
most of our discussions are within the framework of Einstein
manifolds~\cite{Besse:1987}. After then, we consider the
possibilities and phenomenologies of various AdS vacua, especially
the properties of domain walls separate them. We also discuss the
possibilities for a non-conformal boundary field theory to hold a
landscape. Along the opposite root, we studied the consequences of
our conjecture, if QCD (as a non-conformal boundary field theory)
holds a landscape of vacua. In this case, different vacua of QCD
should possess different fundamental parameters, such as the quark
masses $m_q$, the running coupling constant $\alpha_S$, or the CP
violating phase $\theta_\mathrm{QCD}$. Another vacuum with
parameters different from ours may be testable; and we estimate this
possibility within several astrophysical environments.

The organization of this paper is as follows. In
Sec.~\ref{sec:Einstein_manifold} we discuss several mathematical and
physical issues that relate to the Einstein manifolds. We consider
the symmetric conditions of the string worldsheet and the dual field
theories, and the properties of wrapped branes and fluxes. We also
consider the stability conditions of vacua topologies. In
Sec.~\ref{sec:holo_phenomenology}, we discuss the theoretical issues
to approach a QCD landscape. We consider the possibilities to break
CFT, the fundamental parameters a vacuum should determine, and the
deduced parameters that may relate to applications/observations. In
Sec.~\ref{sec:astronomy}, we consider how a QCD landscape affects
astrophysical observations. The applications are abundant, but the
studies in our paper are only tentative. We summarize our results in
Sec.~\ref{sec:conclution}. Some mathematical supplements relatively
independent to the main text are gathered in
Appendix~\ref{app:mathematics_addons}; and the validity of the
orders of magnitudes estimations used in this paper, are
reconsidered more carefully in Appendix~\ref{app:order_estimations}.
We gather these materials together, rather than write two separate
papers from either the theoretical or the astrophysical aspects,
because we think neither one alone is enough to make our conjecture
reasonable; however, the two roots can in fact be read separately.
We always denote the indices of the extended dimensions by $\mu,
\nu$, of the compact dimensions by $m, n$, and of the entire target
space by $M, N$. We set $\hbar = k = c = 1$ for simplification
throughout this paper.

\section{Einstein manifolds and beyond\label{sec:Einstein_manifold}}

Unlike the string compactification $\mathrm{Mink}_4 \times
\mathrm{CY}_3$, where both the four-dimensional $(-~+~+~+)$
spacetime and the compactified manifold are Ricci flat, a
holographic theory may enjoy an AdS vacuum (with a dual CFT) or its
generalizations (with other possible dual field theories like QCD).
For the former case, the metric may be described by a product space
$\mathrm{AdS}_5 \times \mathcal{M}_5$ or $\mathrm{AdS}_5 \times
\mathcal{M}_6$, where $\mathcal{M}_q$ is some compact solution of
the field equation, with its dimension depending on whether the
theory is compactified from a ten-dimensional superstring theory or
an eleven-dimensional M-theory. Although different ways are possible
for choosing $\mathcal{M}_q$, mostly we assume it is an Einstein
manifold follows $R_{mn} \propto g_{mn}$ with positive cosmological
constant~\cite{Duff:1986hr}. As a theorem by Myers, positive
curvature Einstein spaces are always compact (see e.g., \S6.51
of~\cite{Besse:1987}).

Generally, it can be thought that the tremendous landscape of the
Calabi-Yau compactification rises from the abundant type of
Calabi-Yau threefold and the not-very-small Betti numbers $b_m$ of a
typical one. Fluxes are quantized in $m$-cycles (which their number
decided by the $b_m$). As moduli (hence the geometry) of the compact
dimensions are stabilized by the fluxes, different choices of the
quantized condition induce different vacua.

\subsection{Einstein manifolds\label{subsec:Einstein_mfd}}

Hence, to ask whether a holographic theory possesses a landscape of
vacua, our questions are as follows: Whether there are abundant
positive curvature Einstein 5- or 6-manifolds with different
topologies? What properties (such as holonomy groups or isometry
groups) do they possess? Can some of them hold not-very-small Betti
numbers $b_i$? And after holding this set of manifolds in hand, the
follow-up things is to filter them by some additional physical
conditions, such as special topological requirement, or stabilities
of the geometry against small fluctuations.

However, we may incapable to give an all-around or up-to-date
discussion for the mathematical aspect of these questions.
Fortunately, some simple mathematical considerations have already
given us some clues and restrictions to these questions.

Most of the time, researches of Einstein manifolds are limited to
the homogeneous spaces. They are diffeomorphic to coset spaces
$\mathcal{M}_q = G/H$, in which the group $G$ acts transitively on
$\mathcal{M}_q$ (hence, it is the subgroup of the full isometry
group), and $H$ is the isotropy subgroup of $G$ at a point in
$\mathcal{M}_q$. If one is restricted to coset spaces, the complete
list of Einstein manifolds is possibly explored. Some of them are
discussed by physicists in the Kaluza-Klein supergravity
background~\cite{Duff:1986hr}. For example, the list of positive
curvature Einstein 5-manifolds are $S^5$, $SU(3)/SO(3)$, $T^{01} =
S^3 \times S^2$, $T^{11}$, and other $T^{pqr} =
T^{pq}/Z_{r}$~\cite{Romans:1984an}; and of Einstein 7-manifolds are
$T^7$, $S^7$, $J^7$ (squashed $S^7$), $M^{pqr}$, $N^{pqr}$,
$Q^{pqr}$, $S^4 \times S^3$, $(SU(3)/SO(3)_\mathrm{max}) \times
S^2$, $SO(5)/SO(3)_\mathrm{max}$, and
$V_{5,2}$~\cite{Castellani:1983yg}. The $\mathcal{M}_6$ cases seem
more complicated, and the already done researches are closely
related to the compactified mechanisms. For some definite
$\mathcal{M}_6$ and their properties,
see~\cite{Gauntlett:2004zh,Martin:2008pf} and references therein.

There are also manifolds which are not coset spaces but we have
systematic ways to study; e.g., the product spaces $\mathrm{K}3
\times T^{q-4}$ with holonomy group $SU(2)$, or the Calabi-Yau
threefolds. However, these two examples are both Ricci flat. In
addition, Calabi-Yau threefolds are preferred in the
$\mathrm{Mink}_4 \times \mathrm{CY}_3$ scenario, because the
holonomy group need to be $SU(3)$. We do not possess any analogous
restrictions at the very beginning to discuss our cases.

The classifications of the holonomy group $\mathrm{Hol}$ or its
restricted analog $\mathrm{Hol}^0$ (for which the loop is
contractible) with homomorphism $\pi_1 \rightarrow \mathrm{Hol} /
\mathrm{Hol}^0$, may be important for the follow-up studies. If
$\mathrm{Hol}^0$ is reducible, we have (at least locally)
$T(\mathcal{M} = \mathcal{M}' \times \mathcal{M}'') = T'\mathcal{M}
\oplus T''\mathcal{M}$, and $\mathrm{Hol}^0(\mathcal{M}) =
\mathrm{Hol}^0 (\mathcal{M}') \cdot \mathrm{Hol}^0 (\mathcal{M}'')$
as a de Rham decomposition~\cite[\S3.2]{Joyce:2000}. Product spaces
like $\mathrm{K}3 \times S^{1}$ or $\mathrm{K}3 \times T^{2}$ are in
that case. While for the irreducible cases, if the Einstein manifold
is symmetric as $G/H$ in the adjoint representation, its holonomy
group $\mathrm{Hol}^0$ is just $H$~\cite[\S3.3]{Joyce:2000}. And if
it is non-symmetric, the Berger classification said that
$\mathrm{Hol}^0 = SU(3)$, $U(3)$, and $SO(6)$ for $\mathcal{M}_6$,
but only $\mathrm{Hol}^0 = SO(5)$ for
$\mathcal{M}_5$~\cite[\S3.4]{Joyce:2000}. $\mathrm{Hol}^0 = U(3)$
gives K\"{a}hler threefolds, while $\mathrm{Hol}^0 = SU(3)$ gives
Calabi-Yau threefolds in the $\mathrm{Mink}_4 \times \mathrm{CY}_3$
compactifications. However, if for some reasons, we need the
holonomy group of $\mathcal{M}_5$ to be smaller than $SO(5)$, we can
directly rule out all the spaces which are not homogeneous.

For the $\mathcal{M}_6$ cases we may, for some reasons, prefer the
six-dimensional manifold to be K\"{a}hler-Einstein. Then, the first
Chern class $c_1$ of it should have a sign. The condition that $c_1$
is larger (smaller) than zero, gives positive (negative) curvature
$\mathcal{M}_6$, and $c_1 = 0$ gives Calabi-Yau threefolds. In
addition, we have a relation
\begin{equation}
    V \cdot s^3 = \frac{(12 \pi)^3}{3!} c_1^3 ~\mbox{,}
\end{equation}
where $s$ is the scalar curvature, and $V$ is the total volume of
the compact manifold $\mathcal{M}_6$~\cite[\S11.5]{Besse:1987}.
However, although a compact complex manifold with $c_1 \leq 0$
always admits a K\"{a}hler-Einstein metric, for $c_1 > 0$ that
statement is false~\cite[\S11.17]{Besse:1987}. Other interesting
theorems include that compact, complex manifolds with $c_1 > 0$
(they include the positive curvature $\mathcal{M}_6$ cases) have no
non-trivial holomorphic $p$-form~\cite[\S11.24]{Besse:1987}, and are
simply connected thus that $\mathrm{Hol} =
\mathrm{Hol}^0$~\cite[\S11.26]{Besse:1987}.

Whether there are really abundant type of positive curvature
$\mathcal{M}_q$? If including the cases that are not homogeneous,
this question is really hard to answer. However, we may have reasons
to believe that they are much rarer than the negative curvature
ones. While it is easy to find negative curvature
K\"{a}hler-Einstein manifolds, it is hard to find a positive
curvature one; in addition, the known positive curvature ones are
always associated with some isometry
groups~\cite[\S0.I]{Besse:1987}. After normalizing the total volume,
the scalar curvature of $\mathcal{M}_q$ have an upper bound, which
is achieved by $S^q$~\cite[\S12.61]{Besse:1987}. However, this
restriction may be looser for $\mathcal{M}_5$ than for
$\mathcal{M}_6$. The reason is in the $\mathcal{M}_5$ cases, $s$ can
be arbitrarily close to zero, while in the $\mathcal{M}_6$ cases,
they cannot.

Some other issues of $\mathcal{M}_q$ is related to their Einstein
structure moduli spaces. It is really interesting that the Einstein
structure is rigid (that is, an isolated point of the moduli space)
for negative curvature manifolds, but not rigid for the Ricci flat
ones~\cite[\S12.73]{Besse:1987}. The positive curvature cases, which
we are interested in, are much harder to deal with. However, it
should be really important to handle the moduli space of
$\mathcal{M}_q$, if we want to discuss their landscape. The special
case for the K\"{a}hler-Einstein structure, and also the number of
moduli, is discussed in \S12.98 of~\cite{Besse:1987}.

There are some powerful techniques, such as ``toric
variety''~\cite{Greene:1996cy}, to help us study the topology of
(part or all of) the Calabi-Yau threefolds. It is common for a
$\mathrm{CY}_3$ to possess some not-very-small Betti numbers $b_2$
and $b_3$. For some special cases, the positive curvature
$\mathcal{M}_5$ or $\mathcal{M}_6$ may be studies by the conifold
construction, in which the conifold $C(\mathcal{M}_q)$ can be
studies properly; the Sasaki-Einstein $\mathcal{M}_5$'s are already
very abundant. As discussed in Sec.~\ref{subsubsec:condition_SUSY},
we may not limit ourselves to the conifolds, thus we simply assume
that some of them (especially the non-symmetric ones) also own these
properties. Sometimes to avoid distinguishing the holonomy groups
$\mathrm{Hol}$ and $\mathrm{Hol}^0$, we assume $\mathcal{M}_q$ to be
simply connected, hence $b_1 = 0$; but it is in fact not needed. The
hypothesis does not directly contradict with the mathematical
arguments given in this subsection; and it is trusted, as we also
loose some additional constraints such as complex structure or
K\"{a}hler structure. Of course, complex or K\"{a}hler restraints
rise from some definite physical properties of the traditional
compactification, and our questions also have their own physical
conditions; however, it seems not very easy to give definite
(general) constraints from a physical viewpoint, or at least too
early to give up possibilities for scenarios which do not possess
such constraints.

\subsection{Supersymmetric conditions}

For our scenarios of the holographic landscape to work consistently,
there are several different types of supersymmetric conditions. Some
of them need to be held, or need to be held for definite models, but
some others need to be broken for required properties. In this
section, we discuss them separately.

\subsubsection{Restrictions from worldsheet superconformal field theories\label{subsubsec:restrict_ws_CFT}}

For a consistent ten-dimensional superstring theory, the worldsheet
field theory should possess the superconformal symmetry. Consider
the worldsheet theory as a two-dimensional non-linear
$\sigma$-model. If the target manifold is Hermitian and K\"{a}hler,
the worldsheet theory holds a $\mathcal{N} = 2$ supersymmetry; if it
is hyperk\"{a}hler, the worldsheet theory holds a $\mathcal{N} = 4$
supersymmetry, and vice versa if it is
supersymmetric~\cite{Zumino:1979et,AlvarezGaume:1981hm}. While if it
is Ricci flat and K\"{a}hler, the one-loop $\beta$ function of the
worldsheet theory is zero, regardless of the worldsheet
supersymmetry~\cite{AlvarezGaume:1980dk}. We neglect the multi-loop
correlation of the conformal symmetry in this paper. Absolutely,
$\mathrm{Mink}_4 \times \mathrm{CY}_3$ satisfies all these
requirements. However, the conditions given above are sufficient,
but not necessary.

First, a $\mathcal{N} = 1$ worldsheet supersymmetry is enough for a
consistent superstring theory. In this case, the K\"{a}hler
condition is superseded by the existence of a tensor field $J^M{}_N
\in C^\infty(T \mathcal{M}_{10} \otimes T^{*} \mathcal{M}_{10})$
which satisfies $g_{PQ} J^P{}_M J^Q{}_N = g_{MN}$ and is covariantly
constant~\cite{AlvarezGaume:1981hm}. Notice that if in addition
$J^P{}_M J^M{}_N = - \delta^P{}_N$, the target manifold is
K\"{a}hler, but we do not possess such conditions. To keep the
worldsheet conformal symmetry, we need only $g^{MN} \Gamma^P{}_{MN}
= 0$ beside Ricci flatness~\cite{AlvarezGaume:1980dk}, a weaker
condition compared with K\"{a}hler.

For our case $\mathcal{M}_{10} = \mathrm{AdS}_5 \times
\mathcal{M}_5$, $J^M{}_N|_p$ of some point $p \in \mathcal{M}_{10}$
generally represent as a subgroup of $O(10)$, which is invariant
under the action of $SO(5) \times \mathrm{Hol}^0 (\mathcal{M}_5)$.
If $\mathcal{M}_5$ is irreducible and non-symmetric, $\mathrm{Hol}^0
(\mathcal{M}_5) = SO(5)$. The most simple case of $J^M{}_N$ is
$\delta^M{}_N$. More details are given in Appendix~\ref{subapp:JMN}.

Although the ``classical'' $\mathrm{AdS}_5 \times S^5$ configuration
holds Ricci flatness, it does not satidfy the condition $g^{MN}
\Gamma^P{}_{MN} = 0$. See Appendix~\ref{subapp:gGamma} for the
detail calculations.

However, we may not need to take this argument too seriously. The
first reason is about the applicability of worldsheet field theory,
which is only an effective description in the $g_s \rightarrow 0$
limit. In addition, when we study the theory perturbatively (that is
where the arguments of conformal symmetry come from), we give up all
the heavy degree of freedom in string theory. Even if for the
majority of $\mathrm{AdS}_5 \times \mathcal{M}_5$, perturbative
constrain is not a seriously problem, curvature singularities may
exist in the moduli space. If the manifold approaches these
singularities by some dynamical reasons, massive states may become
massless and ruin the perturbativity~\cite{Seiberg:1994rs}. The
second reason is that the AdS/CFT correspondence in the large-$N$
limit always possess small curvatures. Ten-dimensional supergravity
is suitable in that reason, and configurations such as
$\mathrm{AdS}_5 \times S^5$ are indeed solutions supergravity
equations. As we know seldom about how to do a string theory in some
general manifold, a supergravity argument may already be enough.

\subsubsection{Conditions of the dual field theories\label{subsubsec:condition_SUSY}}

The isometry group $SU(4) \simeq SO(6)$ of the prototype
$\mathrm{AdS}_5 \times S^5$ correspondence~\cite{Maldacena:1997re},
gives the global R-symmetry of the four-dimensional $\mathcal{N} =
4$ $U(N_c)$ SYM theory. Similarly, the isometry group of some
typical $\mathcal{M}_q$ in $\mathrm{AdS}_5 \times \mathcal{M}_q$,
gives the remained supersymmetry of the dual CFT. $\mathcal{N} = 2
\times n_{\mathcal{M}_q}$ for type II superstring theory, and
$\mathcal{N} = n_{\mathcal{M}_q}$ for M-theory, where
$n_{\mathcal{M}_q}$ is the number of Killing spinors in
$\mathcal{M}_q$. However, while even dimensional manifolds
$\mathcal{M}_q$ possess equal solutions for both orientations, when
$q$ is odd, solutions can exist only for one orientation unless for
round $q$-spheres~\cite{Acharya:1998db}. Particularly,
$\mathrm{AdS}_5 \times \mathcal{M}_5$ break half of their
supersymmetry and possess only $\mathcal{N} = n_{\mathcal{M}_5}$
(that is also true for the $\mathrm{AdS}_5 \times S^5$ case, because
the geometry is only constructed asymptotically).

There are already several constructive models to break $\mathcal{N}
= 4$ supersymmetry of the boundary CFT; however, most of them still
possess at least the $\mathcal{N} = 1$ supersymmetry. The most
direct construction is for $S^5 / \Gamma$ and their blow-up
manifolds~\cite{Kachru:1998ys}. $\mathcal{N} = 2$ if $\Gamma \subset
SU(2)$, $\mathcal{N} = 1$ if $\Gamma \subset SU(3)$, and the CFT is
still chiral if $\Gamma$ is a complex subgroup of
$SU(4)$~\cite{Lawrence:1998ja}. Another possibility is the conifold
construction~\cite{Klebanov:1998hh}. While M2/M5/D3-brane solutions
can be described as the interpolation between Minkowski spaces and
$\mathrm{AdS}_p \times S^q$, a general $\mathrm{AdS}_{p+2} \times
\mathcal{M}_q$ can be structured by locating large amounts of branes
at some singularity of a conifold $C(\mathcal{M}_q)$, and described
as the interpolation between $\mathrm{Mink}_{p+1} \times
C(\mathcal{M}_q)$ and $\mathrm{AdS}_{p+2} \times
\mathcal{M}_q$~\cite{Acharya:1998db,Morrison:1998cs}. If the
singularity is Gorenstein canonical type, $\mathcal{M}_q$ is
Einstein restricts the cone $C(\mathcal{M}_q)$ to be Ricci flat, and
the Killing spinors on $\mathcal{M}_q$ is in one-to-one
correspondence with the covariantly constant spinors on
$C(\mathcal{M}_q)$~\cite{Bar:1993}. If $\mathcal{N} \geq 1$ is
needed for the dual CFT, $C(\mathcal{M}_q)$ has to hold some special
holonomy. As the Ricci flatness rule out most homogeneous manifolds,
the Berger's classification mentioned in
Sec.~\ref{subsec:Einstein_mfd} restricts $C(\mathcal{M}_5)$ to
$\mathbb{R}^6$, $\mathrm{K}3 \times \mathbb{R}^2$, and
$\mathrm{CY}_3$, and $C(\mathcal{M}_6)$ to $\mathbb{R}^7$,
$\mathrm{K}3 \times \mathbb{R}^3$, $\mathrm{CY}_3 \times
\mathbb{R}^1$, and the $\mathrm{Spin}(7)$-manifold. Especially, the
horizon geometry of the $\mathrm{CY}_3$ conifolds are
Einstein-Sasaki 5-manifolds, which can be described as some $U(1)$
bundle over the K\"{a}hler-Einstein twofold with the Chern class
$c_1>0$~\cite{Kehagias:1998gn}.

However, the more general field theories we interest in this paper,
such as QCD, do not need to possess any supersymmetry. Hence, they
do not need to (and, they are difficult to) be studied
constructively. The $\mathcal{N} = 0$ supersymmetric condition
generally rules out the possibilities to structure the vacua
algebraic geometrically while conifold construction; in addition,
maybe even the Gorenstein canonical singularities (hence, the Ricci
flatness of the cone $C(\mathcal{M}_q)$) are not needed to locate
the branes. Hence, for some general holographic field theories, the
restriction for the isometry group thereafter the metric is
relatively loose.

\subsection{Fluxes, and the stabilization of moduli and topologies}

The abundant Calabi-Yau vacua rise from different patterns of fluxes
wrapped nontrivially on cycles in $\mathrm{CY}_3$. In
Sec.~\ref{subsec:Einstein_mfd}, we argued the possibilities for
Einstein 5 or 6-manifolds with positive curvature to possess
not-very-small Betti numbers and various cycles. Here, we study the
question, that if that is true, whether a landscape of the
holographic vacua is possible or not.

\subsubsection{Some comparisons about flux compactification\label{subsubsec:flux_comparisons}}

There are several differences/comparisons between the
$\mathrm{Mink}_4 \times \mathrm{CY}_3$ and the $\mathrm{AdS}_{p+2}
\times S^q$ compactifications. The most direct one is that, while
the $\mathrm{AdS}_{p+2} \times S^q$ ones are the Freund-Rubin
type~\cite{Freund:1980xh}, the $\mathrm{Mink}_4 \times
\mathrm{CY}_3$ ones are not. We may describe the Freund-Rubin
mechanism in a more modern way. By the de Rham's theorem, given any
set of integers $\nu_n$, $n = 1, \ldots, b_m$, there exists a closed
$i$-form $\omega$ which satisfied $\int_{c_n} \omega = \nu_n$, where
$c_n$ are the correspondent $m$-cycles. A general orientable compact
manifold $S^q$ or $\mathcal{M}_q$ has $b_q = 1$. For the
Freund-Rubin compactification of $\mathrm{AdS}_5 \times S^5$,
$\omega$ is just the dual Ramond-Ramond (RR)-field strength of the
D3-branes, and we have $\int_{S^5} \ast F_5 = N_c$, with the number
of colors $N_c$ of the gauge group of the dual $U(N_c)$ SYM theory.
The radial stability of the $\mathrm{AdS}_5 \times S^5$
configuration is discussed in e.g.~\cite{Silverstein:2004id}, or in
more detail in Sec.~\ref{subsubsec:stabilities}, with the same
solution as the one calculated from the black $p$-brane
supergravity~\cite{Horowitz:1991cd}. The relation between the
RR-charge $N_c$ and the radius of $\mathrm{AdS}_5$ is $R \propto
N^{1/4}$. In the Calabi-Yau cases, fluxes are compactified in the 2
and 3-cycles of $\mathrm{CY}_3$, which their sources D4/6 (in the
IIA case), D5 (in the IIB case), or NS5-branes looking as
$(2+1)$-dimensional domain walls in
$\mathrm{Mink}_4$~\cite{Taylor:1999ii}.

The second difference is as follows. While the flux quantization in
$\mathrm{AdS}_p \times S^q$ is directly related to the gauge group
of the low-dimensional theories, the branes to construct the
Standard Model in $\mathrm{Mink}_4 \times \mathrm{CY}_3$ seem
irrelevant to the ones induce compactification. That makes the
quanta chosen to fix the vacua really optional for the latter
case~\cite{Bousso:2000xa}. The branes to realize the gauge symmetry
in $\mathrm{Mink}_4 \times \mathrm{CY}_3$ has to be space filling.
As their fluxes have nowhere to go, their numbers (the differences
between branes and anti-branes) are highly constrained by anomalies.
However, since the ``real world'' realized in the
$\mathrm{AdS}_{p+2} \times \mathcal{M}_q$ configurations is some
holographic one, flux can goes along the radial direction of the AdS
space.

The third difference is about the existence of maximum flux quanta,
or the finiteness of the absolute number of flux vacua. While the
vacua in $\mathrm{Mink}_4 \times \mathrm{CY}_3$ compactification are
argued to be generally finite~\cite{Douglas:2003um,Acharya:2006zw},
the RR-charge $N_c$ of $\mathrm{AdS}_5 \times S^5$ can be any
integers; hence, the number of the $\mathrm{AdS}_5 \times S^5$ vacua
is infinite. The reason can be understood as below. The finiteness
of vacua in $\mathrm{Mink}_4 \times \mathrm{CY}_3$, restricts from
the tadpole cancelation of the gravitational Cher-Simons
corrections, rises from some global properties of the Calabi-Yau
manifold. Details for type-IIB constraints are constructed, within
the language of F-theory, as
\begin{equation}
    N_\mathrm{D3} + \frac{1}{(2\pi)^4 \alpha'^2} \int H_3 \wedge F_3
        = \frac{\chi(X_4)}{24} \mbox{,}
        \label{eqn:tadpole_cancelation}
\end{equation}
where $N_\mathrm{D3} \geq 0$ is the number of space filling
D3-branes, and $\chi(X_4)$ is the Euler characteristic of the
corresponding Calabi-Yau fourfold~\cite{Sethi:1996es}. However, for
the $\mathrm{AdS}_{p+2} \times \mathcal{M}_q$ cases which can be
described by the horizon geometry of some conifolds, the
restrictions should be completely local; they do not
exist~\cite{Morrison:1998cs}.

Fourth, the no-go theorem~\cite{Maldacena:2000mw}, thus the
requirement of orientifold planes in $\mathrm{Mink}_4 \times
\mathrm{CY}_3$ compactification, need to be reconsidered in the
$\mathrm{AdS}_{p+2} \times \mathcal{M}_q$ cases. Within some
definite assumptions, this theorem ensures that orientifold planes
are needed in $\mathrm{CY}_3$ to get flat or de-Sitter (dS) vacua,
if nontrivial background pattern of fluxes exist. However, since we
are mainly interested in AdS vacuum, orientifold constructions are
not as important as in the string compactification cases.

\subsubsection{Various AdS vacua?\label{subsubsec:various_vacua}}

We argue some landscape of the holographic vacua by the following
reason. Superstring theory has no free parameters. If some more
``realistic'' field theory indeed has its (exact) gravity dual, all
its parameters should be fixed by the moduli of its vacua.
$\mathrm{AdS}_5 \times S^5$, or nearly all constructive vacua
discussed in Sec.~\ref{subsubsec:condition_SUSY}, seem too simple to
accommodate so many parameters.

It is absolutely true that $q$-form flux cycled on $\mathrm{AdS}_5
\times \mathcal{M}_q$, say, the pure Freund-Rubin compactification,
is not enough. As $b_q = 1$ for oriented $\mathcal{M}_q$ and $0$ for
the non-oriented ones, there is only one free quantum to adjust.
Within no \emph{a priori} restrictions about the Betti numbers of
$\mathcal{M}_q$, we may have $F_2$, $F_4$, and $H_3$ fluxes cycle on
$\mathcal{M}_5$ in type-IIA theory, $F_1$, $F_3$, $F_5$, and $H_3$
fluxes cycle on $\mathcal{M}_5$ in type-IIB theory, and $F_4$ flux
cycles on $\mathcal{M}_6$ in M-theory. As $b_n = b_{q-n}$ for any
$\mathcal{M}_q$, the branes taken corresponding RR or
Neveu-Schwarz-Neveu-Schwarz (NSNS)-charges, should always fill in
the Poincar\'{e} dual cycles on $\mathcal{M}_q$, and some
$(3+1)$-dimensional domain walls on $\mathrm{AdS}_5$. Branes as
$(2+1)$-dimensional domain walls on $\mathrm{Mink}_4$ locating in
definite radius of $\mathrm{AdS}_5$ are also possible; however, the
flux configurations are more complicated. To avoid the flux
violating the Lorentz invariance of the dual field theory on
$\mathrm{Mink}_4$, the domain walls should not possess less
dimensions. Similar brane configurations for superstring theory or
M-theory, are described in~\cite{Taylor:1999ii,Gukov:1999ya}.
D5-branes wrapped on two-spheres of AdS vacua, has ever been
discussed for the purpose of broken conformal symmetry (see
Sec.~\ref{subsec:beyond_AdS}, or the review
article~\cite{Bigazzi:2003ui} and references therein). The domain
walls themselves indeed violate the Lorentz symmetry; however, they
can be sat at infinity if needed. The additional requirements may
also include the supersymmetric condition of $\mathcal{M}_q$ cycles;
D-brane instanton or the D-brane spatial components wrapped on them
should possess a supersymmetric worldvolume. The corresponding
Calabi-Yau cycles are considered in~\cite{Bershadsky:1995qy}, in
which ``twist'' is needed. However, we cease for more in-depth
discussions for our case in this work, and leave the relevant issues
to the follow-up studies.

While it is interesting to study flux compactification of some
holographic theory, it is quite difficult to go along a constructive
way, because central charge the dual CFT depends on the fluxes
rather complicated. The first exploration is a type-IIA
compactification on $T^6$ orientifold for $\mathrm{AdS}_4$
vacua~\cite{Aharony:2008wz}. The follow-up works such
as~\cite{Polchinski:2009ch} are also relevant.

In~\cite{Aharony:2008wz}, D4-branes carries RR-charges wrap some
2-cycles on the compact $T^6$ orientifold, and fill their $(2+1)$
other dimensions at some fixed radial position of $\mathrm{AdS}_4$.
However, other configurations of D4-branes are also possible; for
example, the radial $\mathrm{AdS}_3$ inside of $\mathrm{AdS}_4$. We
will ignore the slant configurations in our discussions; they seem
strange when considering the UV-IR
correspondence~\cite{Susskind:1998dq}, as their positions change
while adjusting the energy scale of the dual CFT.

In fact, the latter configuration (in which the branes look as
domain walls in the flat boundary theory) may be more natural. In
our $\mathrm{AdS}_5 \times \mathcal{M}_q$ cases, they are
$(2+1)$-dimensional domain walls within $\mathrm{Mink}_4$, or
$(3+1)$-dimensional domain walls filling in the radial
$\mathrm{AdS}_4$ inside $\mathrm{AdS}_5$. They can be shown from
conifold construction; the analogous M-theory case is described
in~\cite{Gukov:1999ya}. Let us set D7-branes on F-theory background
$\mathrm{Mink}_4 \times C(\mathcal{M}_7)$; D7-branes look as
$(2+1)$-dimensional domain walls on $\mathrm{Mink}_4$, and wrap
5-cycles on 8-conifold $C(\mathcal{M}_7)$. The dual fluxes of
D7-branes wrap 3-cycles on $C(\mathcal{M}_7)$. While locating a very
large number of D3-branes on $C(\mathcal{M}_7)$ singularities, the
spacetime deforms to $\mathrm{AdS}_5 \times \mathcal{M}_7$, as
argued in Sec.~\ref{subsubsec:condition_SUSY}. After then, we can
compact two-dimensions of $\mathcal{M}_7$, to get a type-IIB
superstring theory in ten-dimensions. As the $\mathrm{Mink}_4$ and
the brane within it remain unchanged in this compactification,
branes always look as domain walls in the dual CFT.

Hence, we conjecture that for some more ``realistic'' boundary
theory, the dual string theory on $\mathrm{AdS}_5 \times
\mathcal{M}_q$ possesses other wrapped fluxes beside the volume form
cycled on $\mathcal{M}_q$; the sources of these fluxes fix as domain
walls in the holographic boundary. While the volume form cycled on
the $q$-cycle directly decides the gauge group, different choices of
these other quanta induce gauge theory with different fundamental
parameters.

Are these choices finite or not? In
Sec.~\ref{subsubsec:flux_comparisons}, we discussed the finiteness
of the permutation of quanta (hence the number of vacua) on
$\mathrm{Mink}_4 \times \mathrm{CY}_3$, and the infiniteness of
possible RR-charges on $\mathrm{AdS}_5 \times S^5$. In addition, for
some loose supersymmetric conditions given in
Sec.~\ref{subsubsec:condition_SUSY}, unless the $\mathrm{CY}_3$
cases in which at least the toric description of topology is finite,
the number of $\mathcal{M}_q$ with different topology may even
diverge. Here, we assume the topology is fixed by physical reasons,
and focus on the flux configuration. We guess that the choices for
quanta deciding the fundamental parameters of a holographic theory,
is finite if the corresponding branes display as domain walls
filling radial $\mathrm{AdS}_4$, but infinite if they are restricted
in $\mathrm{Mink}_4$. We leave the strict proof (if exists) to the
follow-up studies, and give an argument as below; the proof is
absolutely complicated, as the geometry is difficult to handle.

The tadpole cancelation, hence, the finiteness described in
Sec.~\ref{subsubsec:flux_comparisons}, can be understand by a finite
energy condition~\cite{Gukov:1999ya}. The energy density, which gets
contribution from both flux and space filling branes (which fill all
uncompact dimensions), should be equal far away on the two sides of
the domain wall. That is true for flat uncompact dimensions, in the
case we show in Eq.(\ref{eqn:tadpole_cancelation}); flux
configuration remain unchanged while going far away from the source.
That is also true for radial $\mathrm{AdS}_4$ within
$\mathrm{AdS}_5$, but not true for the D3-branes laying on
$\mathrm{Mink}_4$ deep inside the throat of $\mathrm{AdS}_5$; in the
latter case, the flux dilutes on the boundary. Things are the same
for the $(2+1)$-filling branes within $\mathrm{Mink}_4$; flux
dilutes on some directions. As these radial $\mathrm{AdS}_4$ and the
$(2+1)$-dimensional domain wall brane configurations should come
together from conifold construction, we argue the holographic vacua
should be infinite.

\subsubsection{Stabilities\label{subsubsec:stabilities}}

Generally, a $\mathrm{AdS}_{p+2}$ vacua is stable, if it satisfies
the Breitenlohner-Freedman bound~\cite{Breitenlohner:1982bm}
\begin{equation}
    m^2 L^2 \geq - \frac{(p+1)^2}{4} \mbox{,}
\end{equation}
where $m$ is the scalar mass of the tachyon mode, and $L$ is the
radiu of the AdS space with the Ricci tensor $R_{\mu\nu} = - (p+1)
g_{\mu\nu}/L^2$. It may be difficult to discuss the stability of the
vacua of some general compact manifold $\mathcal{M}_q$; the mostly
discussed situations are aimed. For the overall Ricci flat
$\mathrm{AdS}_{p+2} \times \mathcal{M}_q$, $\mathcal{M}_q = S^q$ is
always stable, and the Einstein $\mathcal{M}_q = \mathcal{M}_n
\times \mathcal{M}_{q-n}$ for $q < 9$ is always unstable by metric
perturbations~\cite{Ito:1984vi,DeWolfe:2001nz,Shiromizu:2001sp}. For
the free orbifold action $\mathrm{AdS}_5 \times S^5/\mathbb{Z}_k$
with odd $k$ discussed in~\cite{Kachru:1998ys}, while $k = 3$ case
possesses the $\mathcal{N} = 1$ supersymmetry, $k \geq 5$ break all
supersymmetry. The latter cases are unstable~\cite{Horowitz:2007pr}.
In addition, Ref.~\cite{Martin:2008pf} gives the stable condition
for 4-form flux compactified on $\mathrm{AdS}_5$ cross some
$\mathcal{N} = 0$ positive K\"{a}hler-Einstein $\mathcal{M}_6$; the
discussion focuses on homogeneous spaces.

On the whole, supersymmetric conditions can help stabilize the
vacua, as they give some additional restrictions~\cite{Ito:1984vi};
however, they are not absolutely needed. Maybe the discussions of
the $\mathcal{N} = 0$ holographic solutions in
Sec.~\ref{subsubsec:condition_SUSY}, are dangerous. Or maybe a
better way to construct a more ``realistic'' holographic theory, is
to start with AdS vacua with no less than $\mathcal{N} = 1$
supersymmetry, and then break supersymmetry by some other reasons;
as discussed in Sec.~\ref{subsubsec:condition_SUSY}, algebraic
geometrical tools (such as the properties of Einstein-Sasaki
5-manifolds) can be used in that case. It is similar to the idea of
Calabi-Yau phenomenology, where some $\mathcal{N} = 1$ vacua induce
the broken ``real world''. Nonetheless, the relevant issues are
absolutely difficult, as even the stabilization Calabi-Yau vacua is
not easy to handle~\cite{Giddings:2001yu}.

\section{Holographic phenomenologies\label{sec:holo_phenomenology}}

After the more aimed issues related to Einstein manifolds considered
in Sec.~\ref{sec:Einstein_manifold}, we discussed the more
phenomenological aspect of holographic vacua and their landscape,
here in this section.

\subsection{Beyond AdS or beyond $\mathrm{AdS}_5$\label{subsec:beyond_AdS}}

String theory in $\mathrm{AdS}_5 \times \mathcal{M}_5$ should be the
dual theory of some CFTs; the isometry group of $\mathrm{AdS}_5$,
$SO(4,2)$, is just isomorphic to the conformal algebra of the
boundary theory. Hence, to achieve the gravity dual of some more
realistic field theory which is not conformal, we need the
non-compact dimensions deformed from $\mathrm{AdS}_5$.

There are several different ways already discussed, to break
conformal invariance, include: (i) adding a mass deformation to a
CFT, (ii) using wrapped branes -- located on non-vanishing cycles of
$\mathcal{M}_q$, or (iii) fractional branes -- wrapped on collapsed
cycles, and (iv) considering theories at finite temperature.
Generally, additional branes change the blackbrane supergravity
solutions, and finite temperature theories give AdS blackholes
rather than AdS spaces. The first three approaches are reviewed
in~\cite{Bigazzi:2003ui}.

We have already induced wrapped branes in
Sec.~\ref{subsubsec:various_vacua}, for the properties of various
flux compactification patterns. We may generally prefer these
wrapped branes and fluxes induced by them to stabilize the moduli of
$\mathcal{M}_q$, but not alter its geometry (such as its holonomy
group). Maybe a redefinition of the covariant derivative $D_\mu =
\partial_\mu + \omega_\mu \rightarrow \partial_\mu + \omega_\mu +
A_\mu$ is needed, where $\omega_\mu$ is the spin connection, and
$A_\mu$ is the external gauge fields given by the wrapped branes;
the operation is described in~\cite{Bigazzi:2003ui}. This scenario
seems more reasonable in the large-$N_c$ limit, where the D3-branes
Freund-Rubin compactification dominate the geometry. However, things
become specious for some more ``realistic'' theories, such as QCD
which possesses $N_c = 3$. It is difficult to believe the three
D3-branes lead the near-AdS product geometry, while other branes
perturb the exact values of moduli. This may be a general problem
when construction QCD from Large-$N_c$ QCD, which is not easy to
resolve. Similar problem rises from adding flavor branes, which can
only be down in a false assumption -- the probe limit (exact
quenched approximation) $N_f \ll N_c$. For the reason above, we
leave this problem to the follow-up studies.

How should $\mathcal{M}_5$ changes while deforming $\mathrm{AdS}_5$?
If requirements, such as the Ricci flatness discussed in
Sec.~\ref{subsubsec:restrict_ws_CFT}, are needed, they may be
related to each other. Generally, deformations of $\mathrm{AdS}_5$
may ruin the product structure, thus make the definition of a
``vacuum'' ambiguous; however, in the minimum models,
$\mathrm{AdS}_5$ and $\mathcal{M}_5$ may simply be independent with
each other. For example, the temperature of a dual field theory is
always described by the black hole temperature. Take the near
horizon geometry of a black 3-brane solution~\cite{Horowitz:1991cd}
(in this case, the Einstein frame and the string frame are same with
each other)
\begin{equation}
    \mathrm{d}s^2 = H^{-1/2}
        (- f \mathrm{d} t^2 + \mathrm{d} x_1^2 + \mathrm{d} x_2^2
        + \mathrm{d} x_3^2) + H^{1/2} [f^{-1} \mathrm{d} r^2
        + r^2 \mathrm{d} \Omega_5^2(\theta_1, \ldots \theta_5)] \mbox{,}
\end{equation}
where $H = 1 + (R/r)^4$ and $f = 1 - (r_0/r)^4$, we have the
uncompactified dimensions an AdS blackhole solution
\begin{equation}
     \frac{r^2}{R^2}
        (- f \mathrm{d} t^2 + \mathrm{d} x_1^2 + \mathrm{d} x_2^2
        + \mathrm{d} x_3^2) + \frac{R^2}{r^2} f^{-1} \mathrm{d} r^2 \mbox{,}
        \label{eqn:5d_AdS_blackhole}
\end{equation}
with its Ricci scalar $20/R^2$, just like the extreme
$\mathrm{AdS}_5$ case
\begin{equation}
     \frac{r^2}{R^2}
        (- \mathrm{d} t^2 + \mathrm{d} x_1^2 + \mathrm{d} x_2^2
        + \mathrm{d} x_3^2) + \frac{R^2}{r^2} \mathrm{d} r^2 \mbox{.}
        \label{eqn:5d_AdS}
\end{equation}
Notice that the phase transition between
Eq.(\ref{eqn:5d_AdS_blackhole}) and~(\ref{eqn:5d_AdS}), is just the
hard-wall description of the confinement-deconfinement phase
transition~\cite{Erlich:2005qh,DaRold:2005zs}, we argue that a
(Hawking-Page type) phase transition, though maybe changes the
topology of the background spacetime, is possible to be irrelevant
with the compactified dimensions.

\subsection{Fundamental parameters\label{subsec:fundamental_parameters}}

The fundamental parameters of QCD may include: the quark masses
$m_q~(q=u,d,c,s,t,d)$, the coupling constant $\alpha_S$, and the
phase $\theta_\mathrm{QCD}$. If the QCD vacuum corresponds to a dual
gravity theory, all these parameters should be decided by the moduli
of the compact dimensions. To discuss quark masses, one may prefer
to include the Higgs mechanism to the holographic dual; we neglect
the details for these considerations, and simply admit most Standard
Model results if needed. As QCD are always related to other Standard
Model sectors, by the Cabibbo-Kobayashi-Maskawa matrix, or other
scenarios relate the strong, weak, and electromagnetic interactions,
one may prefer for example the $SU(5)$ grand unification theory
rather than the QCD itself, corresponds to some dual gravity theory.
In this case, QCD may rises from some spontaneous symmetry breaking
processes dual to tachyon condensation~\cite{Adams:2001jb}. One may
also prefer to discuss the landscape of some other realistic
systems, such as superfluidity and
superconductivity~\cite{Herzog:2009xv}. We neglect all these
possibilities, and discuss only the landscape of QCD itself here in
this paper.

As the coupling constant $\alpha_s$ is running, it is a little
difficult to consider its relationship with the moduli. In the
perturbative region, one always prefer to describe $\alpha_s (E)$ by
the formula
\begin{equation}
    \alpha_S (E) = - \frac{2 \pi}{(-11 + 2 n_f/3) \ln
    (E/\Lambda_\mathrm{QCD})} \mbox{,}
\end{equation}
where $n_f = 6$ is the number of flavors, and treat
$\Lambda_\mathrm{QCD}$ as a ``fundamental'' parameter. It seems
strange to generalize $\Lambda_\mathrm{QCD}$ to the non-perturbative
regions. Running coupling constants may be understood by the UV-IR
correspondence~\cite{Susskind:1998dq} in the dual gravity theory;
however, quantitative considerations are still difficult.

D-brane physics relates $g_\mathrm{YM}$ and the phase $\theta$ to
the dilaton-axion $\tau$ by
\begin{equation}
    \tau = \frac{4 \pi i}{g_\mathrm{YM}^2} + \frac{\theta}{2 \pi}
        = \frac{i}{g_s} + \frac{\chi}{2 \pi} \mbox{.}
\end{equation}
More specially, the conformal coupling constant of $\mathrm{AdS}_5
\times S^5$ has its relationship with the geometry $g_\mathrm{YM}^2
= 4 \pi g_s = R^4 / \alpha'^2 N_c$, where $R$ is the radius of
$\mathrm{AdS}_5$, and $N_c$ is the number of D3-branes. We may
conjecture in the QCD case that while the dilaton-axion $\tau$
decides $\alpha_S$ and $\theta_\mathrm{QCD}$, other moduli decide
parameters such as quark masses. It seems that the dilaton-axion
should not be constant in the radial direction for a running
$\alpha_S$. We also conjecture that the moduli is decided by the
minimum of some potential $V$ (maybe resemble the one in terms of
the D-terms and the superpotential, used widely in the Calabi-Yau
compactification), and $V_\mathrm{min}$ is to some extent related to
the cosmological constant term hence the radius $R$ of the AdS
space. We may further assume that the relation $4 \pi \alpha_S = R^4
/ \alpha'^2 N_c$ can be generalized (at least in some definite
energy scale/AdS radius) in the $\mathrm{AdS}_5 \times
\mathcal{M}_q$, and also the non-AdS cases discussed in
Sec.~\ref{subsec:beyond_AdS}.

What is the behavior of the domain walls considered in
Sec.~\ref{subsubsec:various_vacua}? The fundamental parameters
possess different values in different sides of the domain wall, as
the moduli do. The potential $V_\mathrm{min}$ should also be
different in each side. Na\"{i}vely, one may thought the domain
walls are infinitely thin, as they are D-branes with codimension
one. However, it is not possible because we deal with quantum
geometry rather than the classical one~\cite{Greene:1996cy}, in
which distances and topologies become meaningless in small scales.
One may turn to think that the domain walls have thickness of the
string scale $l_s$, or the five-dimensional Planck length
$\mathbf{l}_p^3 = l_p^{q+3} / \mbox{volume}(\mathcal{M}_q)$, where
$l_p$ is the Planck length in ten or eleven dimensions. As $l_p =
g_s^{1/4} \alpha'^{1/2} = g_s^{1/4} l_s$, we have $\mathbf{l}_p^3 =
(g_\mathrm{YM}^4 / 16 \pi^2) \cdot l_s^8 /
\mbox{volume}(\mathcal{M}_q)$, and $\mbox{volume}(\mathcal{M}_q)
\sim R^q$. In addition, $\mathbf{l}_p^3 = 15 R^3 / 128 \pi^4 N_c^2$
especially in the $\mathrm{AdS}_5 \times S^5$ case. An alternative
consideration of the domain wall thickness is given in
Appendix~\ref{subapp:ls}.

The metastable vacuum of larger potential should transform/tunnel to
the more stable one. Unlike the bubble nucleation cases, there is no
barrier for this phase transition, and the velocity of the interface
should finally tend to the speed of light $c$~\cite{Bousso:2007gp}.
However, if the potential disparity is really tiny, but the domain
wall is cumbersome, this limit may be difficult to achieve. To
roughly estimate the motion of the domain wall, we assume that it is
at rest in the beginning, and ask its velocity $v$ after time $t$.

All energy rises from the difference of potential $\Delta
V_\mathrm{min}$'s, should transform to the kinetic energy of the
domain wall. As the increase of energy is proportional to the
sweeping distance $s$, the domain wall possesses a constant
acceleration and $s = v t / 2$. By assuming that the domain wall
mass density $\mu_\mathrm{brane}$ is the same order of the tension
of the branes, in the Newtonian limit, energy conservation gives
\begin{equation}
    \Delta V_\mathrm{min} \cdot s = \frac{v^2}{2} \mathbf{T}_3 \mbox{,}
\end{equation}
for some D$p$-brane moving in $(p+1)$-dimensional spacetime, where
the brane tension~\cite{Polchinski:1996na}
\begin{equation}
    \tau_p = \frac{(2 \pi \sqrt{\alpha'})^{1-p}}{2 \pi \alpha' \cdot g_s}
\end{equation}
in string frame and $T_p = g_s^{(3-p)/4} \tau_p$ in Einstein frame.
For branes wrapped on $m$-cycles of compact dimensions, we estimate
the effective tension $\mathbf{T}_{p-m} \sim T_p \cdot
\mbox{volume}(m\mbox{-cycle}) \sim T_p \cdot R^m$. The validity of
the equivalence between $\mu_\mathrm{brane}$ and $\mathbf{T}_3$, is
discussed in Appendix~\ref{subapp:brane_mass_density}.

For branes filling the $\mathrm{AdS}_4$ within $\mathrm{AdS}_5$
discussed in Sec.~\ref{subsubsec:various_vacua}, we have
\begin{equation}
    \Delta V_\mathrm{min} \cdot \frac{vt}{2} = \frac{v^2}{2} \mathbf{T}_3 \sim \frac{v^2}{2}
        \frac{(4 \pi)^{m/4+1}}{(2 \pi)^{m+3}} \frac{N_c^{m/4}}{g_\mathrm{YM}^2 l_s^4} \mbox{,}
\end{equation}
hence
\begin{equation}
    v \sim \frac{(2 \pi)^{m+3}}{(4 \pi)^{m/4+1}}
    \frac{g_\mathrm{YM}^2 l_s^4}{N_c^{m/4}} \Delta V \cdot t \mbox{.}
\end{equation}

How can we estimate $V_\mathrm{min}$ and $\Delta V_\mathrm{min}$?
Absolutely, $V_\mathrm{min}$ has a dominate contribution, rises from
the negative curvature property of $\mathrm{AdS}_5$. If $\Delta
V_\mathrm{min}$ is also in this order of magnitude, we have $\Delta
V_\mathrm{min} \sim 1/R^5 = 1 / N_c^{5/4} g_\mathrm{YM}^{5/2} l_s^5$
and $v \sim (2^{m/2+1} \pi^{3m/4 + 2} / N_c^{(m+5)/4}
g_\mathrm{YM}^{1/2}) (t/l_s)$. Chosen $t \sim 10^{10}\,\mathrm{yr}$
as the age of the Universe, and $l_s = \alpha'^{1/2} \sim
1\,\mathrm{fm}$ as the typical size of a hadron, we have $t/l_s \sim
10^{41}$ and $v \gg 1$. Here, $l_s$ is chosen instinctively from the
Nambu string~\cite{Nambu:1974zg}, or more accurately by the Regge
slope; $\alpha' \sim (1\,\mathrm{GeV})^{-2}$ and $l_s \sim
0.3\,\mathrm{fm}$. The reasonability of the estimations $l_s$ and
$t/l_s$, are discussed in more detail in Appendix~\ref{subapp:ls}.
Hence, the Newtonian approximation break down, and domain walls
should move at the speed of light. However, it is possible that this
contribution of $V_\mathrm{min}$ cancels for different vacua, and
$\Delta V_\mathrm{min}$ rises from other corrections.

One contribution is the intrinsic energy density of the vacuum,
$\rho_\mathrm{vac}$. For the scenario of zero-point energy
fluctuation cutting off at Planck scale, we have $\rho_\mathrm{vac}
= \eta / \mathbf{l}_p^4$ in four dimensions, and $\eta /
\mathbf{l}_p^5$ in five dimensions, where $\mathbf{l}_p$ is the
effective lower dimensional Planck length separately. To solve the
cosmological constant problem~\cite{Weinberg:1988cp}, one need $\eta
= 10^{-120}$ in four dimensions. Assuming $\Delta V_\mathrm{min}$
has the same order of magnitude of $\rho_\mathrm{vac}$, we have
\begin{equation}
    v \sim \frac{2^{m/2+23/3} \pi^{3m/4 + 16/3}}{N_c^{m/4-25/12} g_\mathrm{YM}^{1/2}} \cdot
    \frac{\eta t}{l_s} \mbox{,}
\end{equation}
in the $\mathrm{AdS}_5 \times \mathcal{M}_5$ case. For example, for
branes wrapped on $3$-cycles, $m = 3$ and $v \sim (3.38 \times 10^6
N_c^{4/3} / g_\mathrm{YM}^{1/2}) (\eta t / l_s)$. As $t/l_s \sim
10^{41}$, if $\eta \ll 10^{-47}$, one have $v \ll 1$ and the domain
wall is non-relativistic. It seems possible when comparing with the
cosmological constant case, $\eta \sim 10^{-120}$. However, for the
statistical explanations of tiny
$\eta$~\cite{Bousso:2000xa,Douglas:2003um}, most other vacua possess
$\eta \sim 1$, and a typical $\Delta V_\mathrm{min}$ is not such
small. Deeper reasoning is needed for the more detailed estimation
of $\eta$ in the holographic cases.

Therefore, the domain wall filling the radial $\mathrm{AdS}_4$
within $\mathrm{AdS}_5$ can be either relativistic or
non-relativistic, depends on the magnitude of $\Delta V$; we hold
definite reasons to rule out neither cases. We skip to consider the
other kind of domain walls discussed in
Sec.~\ref{subsubsec:various_vacua}, such as the $(2+1)$-filling
branes within $\mathrm{Mink}_4$; the observational effects and the
dynamical properties seem impalpable in the boundary description.

There is something else to declare, for the non-relativistic branes
discussed above. What should happen if a relativistic particle (for
example, a proton or a neutron) goes across the domain wall? As
discussed above, the fundamental parameters are different in the two
sides; hence, the properties (such as mass or the charge radius) of
the particles should also change. We hypothesize that the energy of
the particles stays the same while passing through the domain wall.

\subsection{Nuclear properties affected by various $\alpha_S$ and $m_q$\label{subsec:nuclear_properties}}

A variation of the QCD coupling constant $\alpha_S$ or quark masses
$m_q$, causes several consequences. Some of them are list as below.

Masses of hadrons and nuclei alter while varying the fundamental
parameters. In the chiral limit where quark and pion masses are
simply neglected, only the change of $\alpha_S$ (or
$\Lambda_\mathrm{QCD}$) plays a role; however, chiral assumption is
not a good assumption for our purpose. The relationship between mass
of pion meson and fundamental parameters, may be estimated by the
Gell-Mann-Oakes-Renner relation~\cite{GellMann:1968rz}. Roughly we
have
\begin{equation}
    m_\pi^2 = \frac{m_u + m_d}{f_\pi^2} \langle 0 | q \bar{q} | 0
    \rangle \sim (m_u + m_d) \Lambda_\mathrm{QCD}\mbox{,}
\end{equation}
as the geometric mean between weak and strong scales; the coupling
of the axial current to pion $f_\pi \sim \Lambda_\mathrm{QCD}$, and
$\langle 0 | q \bar{q} | 0 \rangle \sim
\Lambda_\mathrm{QCD}^3$~\cite{Flambaum:2002de,Kneller:2003xf}.
Masses of protons and neutrons can be alters while strange quark
mass $m_s$ changes, as the strange sea may contribute $1/5$ of the
nucleon mass; however, the dependence of $u$ and $d$ quark masses
are weaker~\cite{Flambaum:2002de,Coc:2006sx}. The strange quark
on-shell mass $m_s = 95 \pm 25\,\mathrm{MeV}$ is absolutely cannot
be neglected.

As the nucleon masses alter, the proton-neutron mass difference and
neutron lifetime also changes. $m_n - m_p$ can be approximately
estimated by
\begin{equation}
    m_n - m_p = m_d - m_u - \xi \alpha \Lambda_\mathrm{QCD}\mathrm{,}
\end{equation}
where $\alpha$ is the fine structure constant, and $\xi$ is a free
parameter which satisfies $\xi \alpha \Lambda_\mathrm{QCD} =
0.76\,\mathrm{MeV}$ at present~\cite{Gasser:1982ap,Kneller:2003xf}.
The neutron lifetime $\tau_n$ depend on this difference by
\begin{equation}
    \frac{1}{\tau_n} = \frac{1}{60} \frac{1+3 g_A^2}{2 \pi^3} G_F^2
    m_e^5 \left[ \sqrt{q^2-1} (2 q^4 - 9 q^2 - 8) + 15 \ln\left(q + \sqrt{q^2-1} \right)
    \right] \mbox{,}
\end{equation}
where $q = (m_n - m_p)/m_e$, and $G_F$ is the Fermi
constant~\cite{Coc:2006sx}. It may happen in some cases, where
neutron is in fact stable.

A variation of the neutron lifetime $\tau_n$, is related to a
variation of the $n \rightarrow p + e^- + \bar{\nu}_e$ reaction
rate. Similarly, the reaction rates of $p + n \rightarrow
{}^2\mathrm{D} + \gamma$ and ${}^2\mathrm{D} + {}^2\mathrm{D}
\rightarrow {}^3\mathrm{T} + p$ are also
changed~\cite{Kneller:2003xf}.

In addition, the stabilities of light nuclei alter. Intuitively
deuterons tend to unbind while $\alpha_S$ (thus also
$\Lambda_\mathrm{QCD}$) decreases; dineutrons and diprotons tend to
be stable while $\alpha_S$ increases~\cite{Barrow:1987sr}. However,
for a quantitative estimation, the critical parameter to control the
nuclear binding energies dominated by pion exchange is $c \equiv
\sqrt{(m_u + m_d)/\Lambda_\mathrm{QCD}}$~\cite{Dent:2001ga}. If for
some definite $c$, the binding energy $E_B < 0$, the correspondent
nucleus is unstable. Deuterons becomes unstable if $c$ decreases by
a factor of $0.77$. Deneutrons become stable if $c$ increases by
$2.6$, while deprotons becomes stable if it increases by $3.2$.
Nevertheless, a first principle estimation of $E_B$ is still
lacking. By assuming a constant $\Lambda_\mathrm{QCD}$, the
variation $\delta E_B / E_B$ may depend on $\sigma$, $\omega$-mesons
and nucleon mass changes separately by contributions proportional to
$\delta m_h / m_h (h = \sigma, \omega, N)$~\cite{Flambaum:2002wq}.

The stabilities of high-Z nuclei are also relevant to
$\alpha_S$~\cite{Broulik:1971vq}. By precondition the liquid drop
model, the stabilized condition is
\begin{equation}
    \frac{Z^2}{A} < \frac{4 \pi r_0^3}{3 e^2} T\mbox{,}
\end{equation}
where $A$ is the atomic number, $Z e$ is the charge, $r_0 \sim
10^{-13}\,\mathrm{cm}$ is the nuclear radius, and $T$ is the surface
tension of the nucleus; in a first approximation we may assume $T
\propto g_\mathrm{YM}^2 = 4 \pi \alpha_S$. Hence unstable nuclei
become stable while $\alpha_S$ increases, and stable nuclei become
unstable while $\alpha_S$ decreases.

Others also argue that the variation of $\alpha_S$ is related to the
single-particle resonance shift $\Delta E_0$ by $\Delta E_r/V_0
\simeq \Delta \alpha_S/\alpha_S$, where $V_0 \sim 50\,\mathrm{MeV}$
denotes the depth of the nuclear potential
well~\cite{Shlyakhter:1976}. Or the proton gyromagnetic ratios $g_p$
depends on fundamental parameters by
\begin{equation}
    g_p = g_p (m_q=0) \left(1 + \sum_q \zeta_q
    \frac{m_q}{\Lambda_\mathrm{QCD}} \right)\mbox{,}
\end{equation}
where $\zeta_q$ is free parameters to denote that this equation is
only a linear approximation~\cite{Flambaum:2002de}.

\section{Astrophysical applications\label{sec:astronomy}}

The AdS/CFT phenomenology in the astrophysical context is only at
its infancy. The influences of AdS/CFT to the cosmological QCD phase
transition are discussed in~\cite{Qiu:2008xc}; and a relation
between the strange quark stars and the Kovtun-Son-Starinets bound,
a direct result from the finite temperature AdS/CFT, is discussed
in~\cite{Bagchi:2007ir}.

In Sec.~\ref{sec:Einstein_manifold}
and~\ref{sec:holo_phenomenology}, we are engaged in a top-down
scenario to discuss whether a holographic theory can possess a
landscape of vacua. That scenario, though exciting, is at most a
conjecture with a huge number of logical and technical
uncertainties. In this section, we try to give a bottom-up argument
of how can a holographic field theory (especially QCD) with
divergent landscape affects our real world. Does it have some
observable applications? Can it solve definite
experimental/observational problems? Different sorts of constraints
of fundamental parameters are reviewed in~\cite{Uzan:2002vq}.
Astrophysical environments have their own advantages for these
questions, as they possess large spatial scales, which may include
domain walls; most terrestrial experiments can only constrain the
variation of the fundamental constants within some definite
timescale. Notwithstanding, maybe a better background to discuss
this problem, is within the areas of nuclear/RHIC
physics~\cite{Mateos:2007ay}, or condensed matter
systems~\cite{Herzog:2009xv}, in which the AdS/CFT phenomenologies
are studies more deeply. Because of the professional background of
the authors, we limit our discussions in the context of
astrophysics, and leave the relevant issues list above to the
follow-up studies.

Unlike the ``multiverse'' discussions caused by Calabi-Yau-kind
landscape, which mainly focus on various gravity-related parameters
such as the cosmological
constant~\cite{Bousso:2008bu,Bousso:2009gx}, the applications of the
AdS/CFT (or simply QCD) landscape seem more abundant. However, we
flung off here only some superficial arguments considered within few
possible environments. Hopefully, more all-around and deep-inside
discussions will come soon.

\subsection{What can we predict?\label{subsubsec:prediction}}

Most former constraints of fundamental constants, limit on their
variations within some definite timescale. Moreover, mostly, authors
assume that they vary smoothly. For our purpose, the minima of vacua
expectation value are fixed, for some definite topology of the extra
dimensions, and quantum numbers of fluxes. Thus, the expected values
of the fundamental parameters are also explicit. By admitting these
preconditions, several phenomenological consequences are possible:

Firstly, smoothly varied constants while time elapses are still
possible. Although the minima are definite, vacua may only tend to
it, and that may be a long-term process. This idea applies widely in
string inflationary models, and should also be practicable in late
universe. The special condition that all parameters are dependent on
a single dilaton field, and the correspondent constraint from
big-bang nucleosynthesis (BBN), are discussed
in~\cite{Campbell:1994bf,Ichikawa:2002bt}. However, as mostly for
this case, the discussions are similar to what given
in~\cite{Uzan:2002vq} and references therein, we neglect to discuss
their consequences here.

Secondly, even if all vacua are in their minima, fundamental
constants in local universe can also change within some timescale.
They change discontinuously. That happens, if a domain wall with its
dynamical properties discussed in
Sec.~\ref{subsec:fundamental_parameters} sweeps us in some definite
ancient epoch. The domain wall can be either relativistic, or
non-relativistic. To clarify, we roughly distinguish two kind of
restraints: (i) The ones focus on local changes of parameters, such
as BBN predictions, stabilities of nuclei, or some other terrestrial
experiments. (ii) The ones focus on far away objects such as pulsars
or quasars, and the conformability of their observational properties
with models. In the latter case, time elapsing is reconstructed by
the long conveyance of photons. Formerly, both kinds of constraints
preconceive a spatial-independent but time-dependent variation. In
our situation, non-relativistic domain walls are suitable for both
restraints; however, relativistic ones cannot be restrained by the
second kind of scenarios. In that case, the other side of the domain
wall is always out of our observable universe. In addition,
non-relativistic domain walls are in some sense difficult to
understand. According to the discussions of
Sec.~\ref{subsec:fundamental_parameters}, the controlled parameter
$\eta$ need to be fine-tuned, to avoid the velocity to be too small;
in this case, the relevant domain walls seem too nearby.

Thirdly, the fundamental parameters may take different values in
different part of the universe, and the (non-relativistic) domain
walls separate them. Few former constraints focus on this
possibility; local restraints are mostly invalid, as the domain
walls are nearly stationary. The consistent conditions needed, and
also the consequences of this possibility, is the major point we
discuss in this section.

\subsection{A consistent condition\label{subsec:domain_wall_consistent}}

Generally, domain walls are precluded in the universe, because their
total masses easily dominate over the matter and radiation
densities. Notice that the energy density is proportional to
$a(t)^{0}$, $a(t)^{-1}$, $a(t)^{-2}$, $a(t)^{-3}$, and $a(t)^{-4}$
for the cosmological constant, domain walls, cosmic strings, matter
(non-relativistic point particles), and radiation, where $a(t)$ is
the scale factor, domain walls easily dominate when $a(t)$ becomes
large.

For our purpose, we need the energy density of the domain walls to
be subordinate to the critical density $\rho_c = 3 H^2 / 8 \pi G
\simeq 1.03 \times 10^{-26}\,\mathrm{kg \cdot m^{-3}}$ (by choosing
the Hubble constant $H = 74.2\,\mathrm{kg \cdot s^{-1} Mpc^{-1}}$).
However, the brane tension
\begin{equation}
    \mathbf{T}_3 \sim
        \frac{(4 \pi)^{m/4+1}}{(2 \pi)^{m+3}}
        \frac{N_c^{m/4}}{g_\mathrm{YM}^2} \frac{1}{l_s^4}
        \label{eqn:T3_tension}
\end{equation}
given in Sec.~\ref{subsec:fundamental_parameters} is for
$(3+1)$-dimensional branes in $\mathrm{AdS}_5$. We do not know how
to calculate the $(2+1)$-dimensional ``holographic'' tension. To
give the right dimensions, two possible choices are $T_\mathrm{dw} =
\mathbf{T}_3^{3/4}$ and $\mathbf{T}_3 \mathbf{l}_p$, where
$\mathbf{l}_p$ is the five-dimensional Planck length; the validity
of the estimation of $T_\mathrm{dw}$, is discussed in
Appendix~\ref{subapp:holographic_tension}.

For domain walls separated by a typical distance $d_\mathrm{dw}$,
their contribution to the energy density is around $\rho_\mathrm{dw}
= T_\mathrm{dw}/d_\mathrm{dw}$. However, for the choices of
$T_\mathrm{dw}$ listed above, $d_\mathrm{dw}$ always seems too
large. For example, by assuming $l_s \sim 1\,\mathrm{fm}$ as the
typical nuclear size, for the case $T_\mathrm{dw} =
\mathbf{T}_3^{3/4} \doteq \kappa/l_s^3$,
$T_\mathrm{dw}/d_\mathrm{dw} < \rho_c$ demand $d_\mathrm{dw} > 6.9
\kappa \times 10^6\,\mathrm{Mpc}$. For example, for branes wrapped
on $3$-cycles, $m = 3$ and $d_\mathrm{dw} > 4.9 \times 10^4
N_c^{9/16}/g_\mathrm{YM}^{3/2}\,\mathrm{Mpc}$, which is already
larger than the scale of the visible universe.

Therefore, if our estimation of domain wall tension is reasonable,
to avoid dominating the energy density, the typical distance
$d_\mathrm{dw}$ should be really large; thus different QCD vacua
should not be testable in \emph{visible} universe. However, it is
entirely possible that the ``holographic'' tension in boundary field
theory, should be calculated in other ways. In that case, the
critical density may not be a strict restraint.

\subsection{Cosmic rays}

Cosmic rays travel long distances to earth. If the regions they
travel hold different fundamental parameters comparing with the
local universe, the ``landscape'' of QCD vacua should leave clues at
observatories. As we argued in
Sec.~\ref{subsec:fundamental_parameters}, energy possessed by the
cosmic ray particles remains the same while crossing the domain
wall, although some other parameters change. Generally, to give
meaningful restraints to confirm/rule out the landscape, the size of
the regions (or the typical distances between the domain walls)
should not be too large, otherwise all possible observations come
from the same region; or too small, that the divergence is fully
averaged.

\subsubsection{Protons}

The energy scale of the Greisen-Zatsepin-Kuzmin (GZK)
cutoff~\cite{Greisen:1966jv,Zatsepin:1966jv} may be altered, as the
masses of proton, $\pi$-meson, and the $\Delta^+$ resonance may
change in different regions. For the reaction $p +
\gamma_\mathrm{CMB} \rightarrow \Delta^+ \rightarrow p' + \pi^0$ (or
$n + \pi^+$), four-momentum conservation gives
$E_{\gamma_\mathrm{CMB}} + E_p = E_{\Delta^+}$ and
$\mathbf{p}_{\gamma_\mathrm{CMB}} + \mathbf{p}_p =
\mathbf{p}_{\Delta^+}$ gives $m_p^2 + 2(E_p E_{\gamma_\mathrm{CMB}}
- \mathbf{p}_p \cdot \mathbf{p}_{\gamma_\mathrm{CMB}}) =
m_{\Delta^+}^2 = (m_p + m_{\pi_0})^2$, thus $E_p = (m_{\pi_0}^2 + 2
m_p m_{\pi_0})/4 E_{\gamma_\mathrm{CMB}}$ for head-on collisions.
For the $2.7\,\mathrm{K}$ cosmic microwave background (CMB) photons,
their typical energy $E_{\gamma_\mathrm{CMB}}$ is around
$1.1\,\mathrm{meV}$, thus we have the cutoff energy $E_p \sim 6
\times 10^{19}\,\mathrm{eV}$. As we have already observed the almost
isotropic CMB radiation, the variation of $E_{\gamma_\mathrm{CMB}}$
by the ``landscape'' is at most a second-order correction. The
cutoff energy scale $E_p$ should indeed changes if $m_p$ or $m_\pi$
alter.

In current, the existence of GZK cutoff has already been confirmed,
but its quantitative properties still hold several
uncertainties~\cite{Cronin:2007zz,Abbasi:2007sv}. By assuming that
there exist a \emph{sharp} cutoff accurately locates at $E_p$, the
distance of the nearest domain walls should be larger than the mean
free path of particles with energy a little above $E_p$. The
required distance is roughly
$100\,\mathrm{Mpc}$~\cite{Aharonian:1994nn}. However, the existence
of a sharp cutoff may be a too strong condition, as a typical
ultra-high-energy cosmic ray (UHECR) source seems within the
$100\,\mathrm{Mpc}$ distance~\cite{Cronin:2007zz}, and the colliding
angles posit randomly. While loosing this requirement, GZK cutoff is
no longer a strict restraint, as a variation of $m_p$ or $m_\pi$ can
at most alters $E_p$ with one order of magnitude.

If the UHECRs in fact come from further sources, the scale of the
regions can still be only around $100\,\mathrm{Mpc}$ if a sharp
cutoff exists. Nevertheless, additional selection principles should
be required, to ensure that the local ``vacuum'' possesses the
smallest $E_p$ comparing with its adjacent regions. The other
possibility is that a variation of $m_p$ or $m_\pi$ also alters the
cross section of $p + \gamma_\mathrm{CMB} \rightarrow \Delta^+$;
however, careful calculations are needed to give further
estimations.

\subsubsection{Neutrons}

Although mostly being neglected, an alternative probability is that
UHECRs are in fact neutrons; air shower experiments such as Pierre
Auger or ARGO-YBJ cannot distinguish protons and neutrons. One skip
this possibility for several reasons: Firstly, as its mean lifetime
is only $885.7\,\mathrm{s}$, an GZK neutron can only travels
$550\,\mathrm{kpc}$ (by chosen $E_p = 6 \times
10^{19}\,\mathrm{eV}$), which is much smaller than the distance of a
mainstream proton source. Secondly, as the Fermi acceleration
mechanism of cosmic rays is an electromagnetic
phenomenon~\cite{Drury:1983}, a neutral particle is hard to
accelerate within it. We reconsider this possibility here, because a
different QCD vacuum may alter the neutron lifetime $\tau_n$, as
discussed in Sec.~\ref{subsec:nuclear_properties}.

As neutrons cannot be influenced by the intergalactic magnetic
fields, one may think this can help explaining the isotropy of
cosmic rays. However, it is unlikely to be so. First, selection
principles are needed to guarantee that the local vacuum possesses
the smallest $\tau_n$, while in other parts of the universe neutrons
hold longevity. Second, it is hard to understand why neutron
composition surpasses proton, if we believe some alternative cosmic
ray producing mechanisms, such as the decay of heavy particles.
Third, one should explain why UHECRs seem coincident with the
supergalactic plane~\cite{Cronin:2007zz}.

\subsubsection{An alternative\label{subsubsec:CR_alternative}}

An alternative possibility is that we indeed live within a domain
wall. This scenario seems too ambitious. However, as we still lack a
reasonable estimation of domain wall thickness, and we have some
ways to wider it, as discussed in Appendix~\ref{subapp:ls}, we
cannot rule it out intuitively. If its thickness is of order the
string length $l_s \sim 1\,\mathrm{fm}$, or the five-dimensional
Planck length $\mathbf{l}_p$, this case may not happen.

One clue is that we all live within the supergalactic plane. In the
mainstream models, large-scale structures like filaments (planes),
haloes or voids, are understood as direct consequences of nonlinear
gravitational effect, and they have already been resulted from
N-body simulations. The coincidence of UHECRs with the supergalactic
plane is easily explained; mass collapse in the structure formation
produces objects such as active galactic nuclei (AGNs) and gamma-ray
bursts (GRBs), and the latter ones are thought to be the sources of
UHECRs. In our scenario, a proper vacuum is a one where the
potential $V$ get its minimum $V_\mathrm{min}$, and a domain wall is
a region where two proper vacua conjuncts (maybe some smoothness
conditions, or the ``domain wall visualizing'' methods, make it
really wide); hence it possesses some large $V$, and some different
vacua. The angle distribution of UHECR patterns may be understood,
if in the normal vacua located at $V_\mathrm{min}$, the fundamental
parameters are disadvantageous for UHECRs to transport; for example,
maybe the mean free path of the reaction $p + \gamma_\mathrm{CMB}
\rightarrow \Delta^+ \rightarrow p' + \pi^0$ or $p'' + e^+ + e^-$ in
these vacua is very small. The $e^+ e^-$ pair production case is
interesting, as its cutoff energy is only $4 \times
10^{17}\,\mathrm{eV}$ in our vacua; however, its cross section is
too low to be considered. If generally in the bulk, the $e^+ e^-$
cross section is larger, UHECRs can only come from direction within
the domain wall. In this case, stellar originations of UHECRs are
not insisted on. As we always lack of plausible ways to observe
these regions, we never know their physics and fundament parameters.
Additional considerations should be needed, to explain why stars or
galaxies never appear in the vacua close to $V_\mathrm{min}$; some
tentative discussions are given in
Sec.~\ref{subsubsec:stellar_evolution}. Despite the fact that our
scenario has nothing to do with the galaxy rotation curves, it may
even explain the dark matter puzzle. If the particles (for examples,
protons or neutrons) in the voids of $V_\mathrm{min}$ are heavier
than which within the domain walls, they can contribute the
additional ``dark'' masses.

\subsection{Abundances of low-Z nuclei}

As already been discussed in Sec.~\ref{subsec:nuclear_properties}, a
different QCD vacuum may possess different hadronic masses,
different neutron lifetime, different reaction rates, or different
binding energies of light nuclei. Here we discuss how they affect
astrophysical observations.

\subsubsection{Big-bang nucleosynthesis}

BBN gives maybe the tightest bound for variations of fundamental
parameters. Starting from~\cite{Barrow:1987sr}, several works focus
on the question of in what regions of variations of
$\Lambda_\mathrm{QCD}$ or quark masses $m_q$, can BBN give a
consistent result with observations. Fundamental parameters mainly
affect BBN through (i) the deuteron binding energy and (ii) the
neutron-proton mass difference. A comprehensive discussion of the
dependence of several deduced and fundamental parameters is given
in~\cite{Dent:2007zu}.

The observation of the primordial abundances of several light
elements, constrain the parameters of BBN. These elements mainly
include ${}^2\mathrm{D}$, ${}^3\mathrm{He}$, ${}^4\mathrm{He}$, and
${}^7\mathrm{Li}$. Most measurements aim to objects within the solar
system, such as meteors, lunar soil, the atmosphere of Jupiter, or
the local universe, such as the interstellar medium (ISM), the Pop I
stars, and the galactic and extragalactic HII regions, which hold
little use for our purpose. Nevertheless, the abundance of deuterium
${}^2\mathrm{D}$ can also be measured by the quasar absorption line.
This gives constraints for the fundamental parameters in the BBN
era, at redshift $z \sim 3$. The up-to-date observational results
can be found in~\cite{Dmitriev:2003qq} and references therein.

Generally, the measurement of the primordial ${}^4\mathrm{He}$ mass
fraction $Y_p$, or some other local abundances, seem more accurate
than the quasar observations of ${}^2\mathrm{D} / \mathrm{H}$.
${}^2\mathrm{D} / \mathrm{H}$ fluctuates from about $1.5 \times
10^{-5}$ to $4 \times 10^{-5}$, for several quasars of redshift from
$2$ to $3.5$. In addition, even after including the observational
errors, these measurements are still inconsistent with each other,
and several of them are out of the weighted mean value
${}^2\mathrm{D} / \mathrm{H} = (2.63 \pm 0.31) \times
10^{-5}$~\cite{Dmitriev:2003qq}. One possible explanation is that
only quasars with low metallicity are suitable for this measurement;
however, the residual metal component can still affect the
observational values. The other possibility is that ${}^2\mathrm{D}
/ \mathrm{H}$ indeed fluctuates here and there, which hints that
fundamental parameters hold a landscape in different part of the
universe.

Here we estimate the variation of other parameters, if
${}^2\mathrm{D} / \mathrm{H}$ intrinsically fluctuates. Notice that
the deuterium abundance is extremely sensitive to the nucleon mass
$m_N = (m_p + m_n)/2$, and $\partial \ln ({}^2\mathrm{D} /
\mathrm{H}) / \partial \ln m_N = 3.5$~\cite{Dent:2007zu}. For
${}^2\mathrm{D} / \mathrm{H}$ varies between $(1.5 \sim 4) \times
10^{-5}$, $m_N$ changes from $799.75\,\mathrm{MeV}$ to
$1058.42\,\mathrm{MeV}$, with the local value $938.92\,\mathrm{MeV}$
corresponds to ${}^2\mathrm{D} / \mathrm{H} = 2.63 \times 10^{-5}$.
Similarly, as $\partial \ln ({}^2\mathrm{D} / \mathrm{H}) / \partial
\ln (m_d - m_u) = -2.9$, we have $(m_d - m_u)_\mathrm{max}/(m_d -
m_u)_\mathrm{min} \sim 1.48$; and as $\partial \ln ({}^2\mathrm{D} /
\mathrm{H}) / \partial \ln (m_d + m_u) = 17$, $(m_d +
m_u)_\mathrm{max}/(m_d + m_u)_\mathrm{min} \sim 1.06$. In addition,
we should emphasis that the fluctuations of $m_N$, $m_d - m_u$, and
$m_d + m_u$ are only some upper limits, for some given distribution
of ${}^2\mathrm{D} / \mathrm{H}$. The value of $\partial \ln
(\bullet) / \partial \ln (\bullet)$ calculated in~\cite{Dent:2007zu}
assume that all parameters vary independently, but the observed
${}^2\mathrm{D} / \mathrm{H}$ is the aggregative effect for all kind
of variations.

Moreover, the divergence of ${}^2\mathrm{D} / \mathrm{H}$ given by
the quasar observations, is incapable to give the upper limits of
the variation of fundamental parameters. A selected effect should be
taken into account, that all possible observations aim to regions
where quasars can be produced. It was argued that mostly the
brightest quasars are hosted by the largest galaxies in the early
universe, and they end up today as central galaxies in rich
clusters~\cite{Springel:2005nw}. Even if we assume that the local
environment (such as the value of fundamental parameters) of quasars
is similar to ours, it is still possible that we all posit in
regions where structure formation are easier than others do.

\subsubsection{Stellar evolution\label{subsubsec:stellar_evolution}}

A different QCD vacuum can absolutely alter the stellar evolution
scenario in several ways. However, as most classical theories in
this field rely on numerical methods, which hold several free
parameters, quantitative discussions of our issue may be really
difficult.

The star formation properties, include the shape of the Hayashi
track, should generally be unaltered by another QCD vacuum.
Gravitational collapse process is mainly caused by gravitational and
electromagnetic phenomena (the latter one is the origin of
dissipation), which are independent of the strong interaction.

The burning of stars, in which energy releases by nuclear fusion
reaction, should be affected if the fundamental QCD parameters
change. The observational Hertzsprung-Russell (HR) diagram may not
rule out the existence of these vacua, as its data points are all
sampled from nearby stars. We may firstly assume that the processes
of proton-proton chain and CNO cycle are still the most important
ones. We neglect to discuss the CNO cycle, as the influences of
other QCD vacua are hard to estimate from the fundamental
parameters. In the proton-proton chain reaction, which mostly
dominates in stars with masses lower than about $1.2\,M_\odot$ in
our QCD, is bottlenecked by the ${}^1\mathrm{H} + {}^1\mathrm{H}
\rightarrow {}^2\mathrm{D} + e^+ + \nu_e$ reaction. Notice that the
cross section of this reaction has a factor
\begin{equation}
    f(x) = (x^2 - 1)^{1/2} \left( \frac{x^4}{30} - \frac{3 x^2}{20} - \frac{2}{15} \right)
        + \frac{x}{4} \log\left[ x + (x^2 - 1)^{1/2} \right] \mbox{,}
\end{equation}
where $x = (2 m_p - m_\mathrm{D})/m_e$ and $m_\mathrm{D}$ is the
mass of ${}^2\mathrm{D}$ without electrons~\cite{Bethe:1938yy}, the
reaction rate relies sensitively on the proton and deuterium masses.
If $x$ is smaller in other QCD vacua, the lifetime of low mass main
sequence stars, should be tremendously longer.

It is possible that in some QCD vacua, stars cannot even be ignited,
either because the reaction rates of the proton-proton chain and CNO
cycle are both too small, or because these processes are not
possible for those parameters. The observational effect should be
dark voids. It is not entirely impossible, as ever been roughly
discussed in Sec.~\ref{subsubsec:CR_alternative} in the background
of cosmic rays; however, as an parallel idea confronts the ordinary
structure formation scenario, detailed discussions are needed for
its consistency and reasoning.

The latter burning phases of stars are also altered for different
QCD parameters. We give up the quantitative discussions, as the
feedback is hard to estimate. Nevertheless, there is one possibility
to mention. Several astrophysical events are difficult to comprehend
from theoretical levels; for example, supernovae never explode in
computer simulations. Is it possible that they explode because they
locate in regions with a different QCD? However, this explanation is
generally unlikely to be so, because one know example (SN 1054; the
Crab Nebula) is really nearby.

The final stage of stars is also different for different QCD. In
reality, stars end as white dwarfs, neutron stars or black holes,
with the former two possess some upper mass limits, called the
Chandrasekhar mass limits. For the case of the electron-degenerate
matter, the mass $M_\mathrm{max} \propto (\mu_e m_\mathrm{H})^{-2}$,
where $\mu_e$ is the molecular weight per electron and
$m_\mathrm{H}$ is mass of the hydrogen atom, depends but is not very
sensitive to the QCD parameters.

\subsection{Quark matter}

In our ``real world'', the energy per baryon number of
${}^1\mathrm{H}$, ${}^{12}\mathrm{C}$ and ${}^{56}\mathrm{Fe}$ is
$\mathcal{E}_h = 938.8$, $931.5$, and $930.4\,\mathrm{MeV}$
respectively. In the case of only $u$ and $d$ quarks, from nuclear
observations we know that nuclear matter is absolutely more stable
than quark matter. It was conjectured that the true zero temperature
and pressure ground state of is the ``strange quark matter'', the
quark-gluon plasma (QGP) of $u$, $d$, and $s$
quarks~\cite{Witten:1984rs}. The main reason is that the additional
strange freedom lowers the Fermi energy. In case of the MIT bag
model~\cite{Chodos:1974je,DeGrand:1975cf}, one has $\alpha_S = 0$
and $m_q = 0$ (especially $m_s = 0$). Assuming that the system is
electrically neutral, thus $2 n_u/3 - n_d/3 - n_s/3 - n_e = 0$ and
$n_e \sim 0$; and the pressure $p_F^4/4\pi^2$ is equal for the
$(ud)$ or $(uds)$ QGP, where $p_F = \hbar (3 \pi^2 n_q)^{1/3}$ is
the Fermi momentum. Notice that $p_{F,u} : p_{F,d} = 1 : 2^{1/3}$
for the $(ud)$ case, and $p_{F,u} : p_{F,d} : p_{F,s} = 1 : 1 : 1$
for the $(uds)$ case. The average quark kinetic energy is generally
proportional to $p_{F,q}$ of that particle, thus we have
\begin{equation}
    \frac{\mathcal{E}_{uds}}{\mathcal{E}_{ud}} =
    \frac{\left(\frac{1}{3} + \frac{1}{3} + \frac{1}{3}\right)
        \left( \frac{1 + 2^{4/3}}{1 + 1 + 1} \right)^{1/4}}
        {\left(\frac{1}{3} + \frac{2}{3} \cdot 2^{1/3}\right)}
        \simeq 0.887 \mbox{.}
\end{equation}
Assuming that $\mathcal{E}_{ud} \sim \mathcal{E}_h$, we have $\Delta
\mathcal{E} = \mathcal{E}_h - \mathcal{E}_{uds} \sim
100\,\mathrm{MeV}$. Hence if $m_s \lesssim 100\,\mathrm{MeV}$,
strange quark matter is more stable than hadronic matter. These
matter ground state conjecture is consistent with the nowadays
constraint is $m_s = 95 \pm 25\,\mathrm{MeV}$.

The existence of charm quarks is generally ruled out from the ground
state conjecture. Charm quark seems too heavy, which a typical mass
$m_c = 1.25 \pm 0.05\,\mathrm{GeV}$. As and $p_{F,u} : p_{F,d} :
p_{F,s} : p_{F,c} = 1 : 2^{1/3} : 2^{1/3} : 1$ for the $(udsc)$ QGP,
roughly we have
\begin{equation}
    \frac{\mathcal{E}_{udsc}}{\mathcal{E}_{ud}} =
    \frac{ \left(\frac{1}{6} + \frac{2^{1/3}}{3} + \frac{2^{1/3}}{3} + \frac{1}{6}\right)
        \left( \frac{1 + 2^{4/3}}{1 + 2^{4/3} + 2^{4/3} + 1} \right)^{1/4}}
        {\left(\frac{1}{3} + \frac{2}{3} \cdot 2^{1/3}\right)}
        \simeq 0.810 \mbox{.}
\end{equation}
Thus, we need $m_c + m_s/2 \lesssim 200\,\mathrm{MeV}$ to make the
``charm quark matter'' the true ground state, which is absolutely
impossible.

However, within our discussions of QCD landscape, it is possible
that in some other vacuum $m_c$ is not so heavy, thus the charm
quark matter is in fact more stable. Or maybe in some vacuum,
$\mathcal{E}_{ud} < \mathcal{E}_h$, therefore hadron states cannot
even exist in the zero temperature and pressure case.

\subsubsection{Strange stars or charm stars}

The strange star is a theoretical model of compact star, which
hypothesizes that compact star is composed of strange quark
matter~\cite{Haensel:1986qb,Alcock:1986hz}. It is a direct corollary
of the conjecture that the strange quark matter is the true ground
state of matter. It is argued that the strange star, rather than the
neutron star, is the true origin of pulsar. Both the strange and the
neutron star possess some Chandrasekhar mass limit of $M_\star \sim
1 M_\odot$ at the radius $R_\star \sim 10\,\mathrm{km}$; however,
the smaller the lighter strange stars, but the larger the neutron
ones. In general, one can calculate the strange star mass-radius
relationship, using the Tolman-Oppenheimer-Volkoff equation and some
proper equation of state (EoS), integrating from the center of the
star, and indicating the surface as the radius where $P = 0$.

For some QCD vacuum in which $\mathcal{E}_{udsc} < \mathcal{E}_h$,
the hypothesized star should in turn be the ``charm star''. Charm
stars have ever been discussed in the realistic
QCD~\cite{Kettner:1994zs}, which do not possess stable charm matter.
Hence, in their case, charm stars exist when the central energy
density of the star is really high, and $P = 0$ is no longer a
reasonable assumption. They conclude that charm stars are generally
unstable by perturbations, thus should not exist. In our case, for
vacua where $m_c$ and $\mathcal{E}_{udsc}$ are much smaller, charm
stars can exist even for really small masses and radii.

However, one may feel difficult to discriminate between strange
stars and charm stars far away in the sky. Consider the EoS of the
MIT bag model, hence the mass-radius relationship of the star. In
the first approximation of $m_s = m_c = 0$, this EoS is independent
of the flavor number, and both strange and charm quark matter, or
even the $(ud)$ QGP, hold the EoS of $P = (\rho - 4B) / 3$, where
$B$ is the bag constant. This EoS should generally be suitable even
if $m_q \neq 0$, as when $m_q$ becomes dynamically important, its
abundance decrease. $B$ should absolutely depends on $m_q$ and
$\alpha_S$; however, in a straightforward understanding of the bag
model, $m_q = \alpha_S = 0$ is assumed. Therefore, strange stars and
charm stars possess similar mass-radius relationships; the only
different is that for the charm star cases, the effective bag
constant $B$ is different from what estimated from baryon resonance
states on earth.

The other possibility is that the surface properties are different
for strange and charm stars; however, it is also not likely to be
so. The quark surface should be really thin, which possesses a
typical strong length scale of order $1\,\mathrm{fm}$; however,
electronic distribution should be more diffused. Thereby a strong
electric field exists near the star surface. Both the cases of
strange star surface with~\cite{Alcock:1986hz} or
without~\cite{Usov:1997ff} the hadronic ``crust'', have ever been
considered. Intuitively, charm stars should hold a much stronger
electric field than the strange stars; as $n_u \neq n_d$ and $n_s
\neq n_c$, the overall neutralized condition makes $n_e$ larger in
charm quark matter. In contrast, strange quark matter have $n_u :
n_d : n_s \simeq 1 : 1 : 1$. Detailed calculations show that
$n_e^{(udsc)} \sim 100 n_e^{(uds)}$ for the same baryonic number
density (see Fig.~3 of~\cite{Kettner:1994zs}, for example).
Quantitatively, one always estimates the surface configuration by
the Thomas-Fermi model. Our discussion limits to the bare quark star
case, in which the crust does not exist. While $n = p_F^3 / 3 \pi^2$
for fermions like electrons or quarks, equilibrium condition gives
the chemical potential at infinite $\mu_\infty = p_{F,e} - V_e =
p_{F,q} - V_q = 0$, where $V_e$ ($V_q$) is the electrostatic
potential for electrons (quarks). Poisson equation gives
\begin{equation}
    \frac{d^2 V_e}{d z^2} = \left\{
        \begin{array}{ll}
            4 \alpha (V_e^3 - V_q^3) / 3 \pi & z \leq 0\\
            4 \alpha V_e^3 / 3 \pi & z > 0
        \end{array}
    \right.
    \mbox{,}
\end{equation}
where $\alpha$ is the fine structure constant, and $z$ is the height
measured from the stellar surface, It possess a solution
\begin{equation}
    V_e = \frac{3 V_q}{\sqrt{6 \alpha / \pi} V_q z + 4}
\end{equation}
for $z > 0$, and a electric field $E_e = - d V_e / d z$. Notice that
$V_q$ is only related to the quark matter density, which is
insensitive to $n_e$ or the existence of charm quark, so do $V_e$
and the electric field $E_e$.

\subsubsection{Strangelets or charmlets}

Strangelets have already been discussed widely in the issue of the
cosmological QCD phase transition, in~\cite{Witten:1984rs} and the
follow-up works. They may also be produced in ultra-relativistic
heavy ion collisions. Reference~\cite{SchaffnerBielich:1998ci} gives
a detailed study of charmlets; however, as charm quarks are too
heavy, charmlets exist in reality should always be unstable.
Estimated from perturbative QCD, an attractive force exists for
large enough quark mass $m_q$; the author explain this phenomenon by
a MIT bag model with the Casimir energy of the bag. In some other
vacua, $m_c$ is not so heavy, and charmlets may be stable or at
least easily produced. In the first case, charmlets may be the relic
of the cosmological QCD phase transition.

In general, light charmlets or charm resonance states, even if exist
in nearby vacua, cannot diffuse to our vacua. The reason has already
been discussed in Sec.~\ref{subsec:fundamental_parameters}. When
charmlets go through the domain wall, the fundamental parameters
change, and the nuggets decay to more stable matter.

An interesting thing is that maybe in some vacua, the $(ud)$ QGP is
the true ground state. Generally, in this case, hadronic matter are
not stable, and stars formed by the gravitational collapse cannot
burn. There are two possibilities. On the one hand, maybe at high
temperature and low pressure regions, QGP is also more stable than
hadronic matter. The cosmological QCD phase transition cannot happen
in this case, and the late universe is either full of quark
nuggets/quark stars, or full of black holes. The latter case happens
when single pieces of quark matter leave after the big-bang, have
their masses surpass the Chandrasekhar limit. On the other hand, if
the hadronic matter is more stable at high temperature,
hadronization indeed happens in the early universe. In this case, if
the potential barrier to transform from hadronic matter to quark
matter is low, the transformation goes smoothly while gravitational
collapse, and the final state is small quark nuggets. Larger objects
are not possible, as dissipations do not exist in this system.
However, if the barrier is high, transformation may only happens
within some violent process, as at really high pressure we always
have the $(udscbt)$ QGP the most stable one. As these violent
processes should not happen within our visible universe, this
possibility is excluded. However, it may be a little early to rule
out the other possibilities, that there are regions in the universe
where (small or large) quark pieces floating in the sky. They behave
like dark voids.

\section{Discussions and conclusion\label{sec:conclution}}

In this paper, we discussed the possibilities that whether QCD, or
more generally some holographic $(3+1)$-dimensional gauge theories,
can possess a ``landscape'' of its vacua. We first limited our
discussions to some boundary CFTs, for which the compact manifolds
are Einstein. An Einstein 5- or 6-manifold $\mathcal{M}_q$, which
needs not to be homogeneous, may have some not-very-small Betti
numbers $b_m$. Therefore, if fluxes wrap different patterns on
cycles of it, the moduli are also different; a landscape of the
holographic vacua should rise for this reason. We examined several
relevant issues for this conjecture from the theoretical viewpoint.
The geometry of $\mathrm{AdS}_5 \times \mathcal{M}_q$ should also
possess some symmetric conditions, such as the worldsheet
superconformal symmetry, or definite supersymmetries of the dual
field theories. However, it seems that even the $\mathrm{AdS}_5
\times S^5$ violates the one-loop worldsheet conformal symmetry. The
isometry group of $\mathcal{M}_q$ decides the R-symmetry of the
holographic theory, which is not imposed in our case. In addition,
we considered the anomaly cancelation, the no-go theorem, the
brane/flux configurations, and the vacua stabilities. We focused on
the $(2+1)$-dimensional domain walls within $\mathrm{Mink}_4$, and
also the $(3+1)$-dimensional domain walls filling the radial
$\mathrm{AdS}_4$ inside $\mathrm{AdS}_5$. We considered the
possibilities to break the $\mathrm{AdS}_5$ geometry, or
equivalently the conformal symmetry of the boundary theory, and
argued that the confinement-deconfinement phase transition may not
affect the vacua configurations. After then, we applied our
conjecture of the ``holographic landscape'' directly to QCD, for
which the fundamental parameters such as $m_q$, $\alpha_S$ or
$\theta_\mathrm{QCD}$ should depend on the moduli of the compact
dimensions. We studied the properties of the domain walls, such as
their masses, thicknesses, and their dynamical properties; they may
be both relativistic and non-relativistic. We discussed how this
``QCD landscape'' affects nuclear physics; they may alter the
hadronic masses, the reaction rates, and the stabilities of the
nuclei.

In an opposite way, we studied how can a ``QCD landscape'' affects
the astrophysical observations. As domain walls may be
non-relativistic, another vacuum of QCD may be within the visible
universe; if they are not, we can also consider the properties of
the other multiverses of QCD, just as what is done
in~\cite{Bousso:2008bu,Bousso:2009gx}. We first considered whether
the mass contribution of the domain walls exceeds the critical
density of the universe. Most of the case, domain walls are really
dangerous; however, it is not enough reasonable that they should be
completely ruled out. We then discussed the properties of cosmic
rays affecting by this landscape, include the GZK cutoff, the
neutron lifetime, and an alternative explanation of the coincidence
of the cosmic ray anisotropic spatial distribution with the
supergalactic plane. The GZK cutoff depends on, but is not very
sensitive to, a different QCD vacua; the variation of the neutron
lifetime seems helpless to explain the observations. The alternative
explanation may loose the constraints for the origins of the UHECRs;
however, as too ambitious the scenario is, it needs to be studied
more seriously. We also considered how the QCD landscape affects BBN
and the star burning scenarios. As only the abundance of deuterium
${}^2\mathrm{D}$ can be measured in far away part of the universe,
only it can restrain the QCD landscape; the constraint is really
loose. A different QCD vacuum can also affect the stellar evolution;
for example, the time spent for the proton-proton chain reaction
depends on the deuterium mass $m_\mathrm{D}$ very sensitively. The
Chandrasekhar mass limits of white dwarfs and neutron stars are also
altered, but not very sensitive to the QCD landscape. In addition,
whereas the ``strange quark matter'' may be the true matter ground
state of our QCD~\cite{Witten:1984rs}, charm or $(ud)$ QGP may be
the ground states of other QCDs; thus, as an alternative, charm
stars or $(ud)$ quark stars may exist in those QCDs. However, the
mass-radius relationships and the surface properties of those
objects seem really similar to our strange stars.

Recently, Denef and Hartnoll discussed the landscape in condensed
matter systems, which results a statistical distribution of critical
superconducting temperatures~\cite{Denef:2009tp}. In here, we
compare briefly the original ``string landscape'', their ``atomic
landscape'', and our ``QCD landscape''. First, whereas the
relativistic quantum critical theories are conformal, our QCD
landscape should break conformal symmetry by some mechanics. Second,
the original Calabi-Yau compactification possesses a $\mathcal{N} =
1$ supersymmetry, the ``atomic landscape'' in M-theory by the
Sasaki-Einstein 7-manifolds holds a $\mathcal{N} = 2$ supersymmetry;
however, it seems that our ``QCD landscape'' is not restricted by
supersymmetry conditions. Third, as the ``atomic landscape''
discussed in~\cite{Denef:2009tp} is limited to the Freund-Rubin
compactification, only the background electromagnetic four-flux
contributes to the vacua field equations; thus their statistics of
the landscape rises from the different topologies of the compact
dimensions. However, the original Calabi-Yau vacua, or the QCD vacua
discussed in this paper, are more abundant because of different flux
configurations.

\section*{Acknowledgements}

I would like to thank Ofer Aharony, Xiao-Jun Bi, Gang Chen, Edna
Cheung, Shamit Kachru, Xin-Lian Luo, Juan Maldacena, Jonathan
Martin, Darren Shih, and Renxin Xu for helps of this work.

\appendix

\section{Mathematical supplements\label{app:mathematics_addons}}

\subsection{About $J^M{}_N$\label{subapp:JMN}}

The holonomy we use in the context is a very special one: the
holonomy of a Riemannian manifold of the Levi-Civita connection on
the tangent bundle of $\mathcal{M}_n$. In this context,
$\mathrm{Hol}(\mathcal{M}_n)$ is a subgroup of $O(n)$.

In our case, $J^M{}_N \in C^\infty(T \mathcal{M}_{10} \otimes T^{*}
\mathcal{M}_{10})$ is a covariantly constant tensor field thus
$\nabla_P J^M{}_N = 0$, to possess worldsheet supersymmetry under
\begin{equation}
\begin{split}
    \delta \varphi^M &= J^M{}_N \overline{\varepsilon} \psi^N\\
    \delta (J^M{}_N \psi^N) &= [-i \dels \varphi^M +
        \frac{1}{2} \Gamma^M{}_{NP} J^N{}_Q J^P{}_R
        (\overline{\psi}^Q \psi^R)] \varepsilon \mbox{.}
\end{split}
\end{equation}
Let $H = \mathrm{Hol}_p(\mathcal{M}_{10})$ be the holonomy group of
a point $p \in \mathcal{M}_{10}$, which acts on a tensor field by
parallel transport its vector bases separately. We denote the
restricted analog of $H$ to be $H^0$. The necessary and sufficient
condition for the existence of a tensor field $J^M{}_N$, is that
$J^M{}_N|_p$ is fixed by the action of $H$ on $T \mathcal{M}_{10}
\otimes T^{*} \mathcal{M}_{10}$~\cite[\S2.5.2]{Joyce:2000}.

Generally, we have $H^0 = SU(5)$, $U(5)$, $SO(10)$ in the Berger
classification. For a manifold which is not complex, the unitary
groups such as $SU(5)$ or $U(5)$ can be generally ruled out. In our
case, the configuration $\mathrm{AdS}_5 \times \mathcal{M}_5$ gives
$H^0 = SO(5) \times \mathrm{Hol}^0(\mathcal{M}_5)$; and if
$\mathcal{M}_5$ is irreducible and non-symmetric, the Berger
classification restricts $\mathrm{Hol}^0(\mathcal{M}_5)$ to be
$SO(5)$.

As the condition $g_{PQ} J^P{}_M J^Q{}_N = g_{MN}$ gives $J J^T =
J^T J = 1$, we have $J \in O(10)$. Hence, a suitable $J^M{}_N$
belongs to a subgroup of $O(10)$, which is invariant under the
action of $H$. We limit our discussions to the simply connected
cases $\mathcal{M}_{10}$ for simplicity; thereafter $H = H^0$. We
also assume $\mathrm{det}(J) = 1$. The most general $H^0 = SO(10)$
constrains $J^M{}_N = \delta^M{}_N$ as the only possibility. The
addition restriction $J^P{}_M J^M{}_N = - \delta^P{}_N$
in~\cite{AlvarezGaume:1981hm} forces $H^0$ to be a subgroup of
$U(5)$, hence gives K\"{a}hler manifolds. In our cases, we restrict
$J$ to be invariant under the action of $SO(5) \times
\mathrm{Hol}^0(\mathcal{M}_5)$.

\subsection{About $g^{MN} \Gamma^P{}_{MN} = 0$ \label{subapp:gGamma}}

Generally, K\"{a}hler manifolds always satisfy $g^{MN}
\Gamma^P{}_{MN} = 0$. The reason is that the only non-zero
components for $g$ is $g_{\alpha \bar{\beta}}$ and $g_{\bar{\alpha}
\beta}$, but the Levi-Civita connection has no mixed indices.

$\mathrm{AdS}_5 \times S^5$ metric does not possess $g^{MN}
\Gamma^P{}_{MN} = 0$, although it holds Ricci flatness. By choosing
the coordinates
\begin{equation}
    \mathrm{d}s^2 = \frac{r^2}{R^2}
        (- \mathrm{d} t^2 + \mathrm{d} x_1^2 + \mathrm{d} x_2^2
        + \mathrm{d} x_3^2) + \frac{R^2}{r^2} \mathrm{d} r^2
        + R^2 \mathrm{d} \Omega_5^2(\theta_1, \ldots \theta_5) \mbox{,}
\end{equation}
we have
\begin{equation}
    g^{MN} \Gamma^P{}_{MN} = \left(
        \begin{array}{l}
            0 \\
            0 \\
            0 \\
            0 \\
            - 5 r / R^2 \\
            4 \sin \theta_1 / R^2 \cos \theta_1 \\
            3 \sin \theta_2 / R^2 \cos^2 \theta_1 \cos \theta_2\\
            2 \sin \theta_3 / R^2 \cos^2 \theta_1 \cos^2 \theta_2 \cos
            \theta_3 \\
            \sin \theta_4 / R^2 \cos^2 \theta_1 \cos^2 \theta_2 \cos^2
            \theta_3 \cos \theta_4 \\
            0
        \end{array}
    \right) \mbox{.}
\end{equation}

\section{Orders of magnitudes reconsidered\label{app:order_estimations}}

\subsection{$l_s \sim 1\,\mathrm{fm}$?\label{subapp:ls}}

$l_s = \alpha'^{1/2} \sim 1\,\mathrm{fm}$ comes from the studies of
the Nambu string~\cite{Nambu:1974zg}, or the Regge slope. It is the
fundamental scale of the QCD string. However, $t / l_s \sim
10^{41}$, which is used in Sec.~\ref{subsec:fundamental_parameters},
may not be reasonable. The reason is that the ``time'' in
$(4+1)$-dimensional gravity corresponds to the boundary QCD, may not
equal to the ``time'' of our $(3+1)$-dimensional gravity.

It was generally said in Sec.~\ref{subsec:fundamental_parameters}
that all free parameters of QCD, such as $m_q$ or $\alpha_S$, are
decided by the moduli of the compact manifold. However, this
judgement may be a little too na\"{i}ve; the quark masses $m_q$, or
the $(3+1)$-dimensional ``time'', are completely the gravitational
effect. Thereafter the question turns to be: how can we induce
gravity in the boundary theory?

One known method to solve this question, is to add the ``Planck
brane'' as some (field) UV
cutoff~\cite{Verlinde:1999fy,Gubser:1999vj,ArkaniHamed:2000ds}. As a
generalization of the Randall-Sundrum (RS) I
model~\cite{Randall:1999ee}, this brane induces gravity, whereas QCD
is located at some other branes in the more IR regions. Unlike the
generalized RS I models, QCD is not fixed at the ``TeV brane'' of RS
I, as ever discussed in~\cite{ArkaniHamed:2000ds}. In this case, the
string scale is of order the $(3+1)$-dimensional Planck length
$l_\mathrm{p,4} = \sqrt{G} = 1.6 \times 10^{-20}\,\mathrm{fm}$ at
the radius of the ``Planck brane'', but of order $1\,\mathrm{fm}$ at
the QCD visualizing radius. However, the time $t \sim
10^{10}\,\mathrm{yr}$ is also measured at the ``Planck brane''
(which is located at the gravity IR, by the UV-IR
relation~\cite{Susskind:2003kw}), and is tremendously redshifted; by
pulling it back to the QCD radius, one has $t \sim
(l_\mathrm{p,4}/1\,\mathrm{fm}) \times 10^{10}\,\mathrm{yr}$ and a
much suppressed $t/l_s \sim 10^{21}$.

Nevertheless, one may think there are some inconsistencies for the
above considerations. By estimating in regions of larger radius, at
the ``Planck brane'' for instance, the value of $t/l_s$ is much
larger. That may in fact hint that for the $\mathrm{AdS}_4$ branes
inside $\mathrm{AdS}_5$, the part away from the throat is more
easily accelerated. However, this distortion of branes is hard to
understand; points located at different radial positions is not
really independent, they are only ``holographic images'' with each
other~\cite{Verlinde:1999fy}. It seems a too tough issue, which we
left to the follow-up studies.

In Sec.~\ref{subsec:fundamental_parameters}, we also argued that the
domain walls may have the thickness of the string scale $l_s$. It is
true if the ``domain wall visualizing'' radius is at the same
position as the QCD brane. However, it may generally not need. Just
as what discussed above, if the ``domain wall visualizing'' radius
is deeply inside the throat of AdS, the string scale there,
thereafter the domain wall thickness, can be much larger. The other
possibility to widen the boundary domain walls, is to consider slant
branes, which we fall to discuss in this paper. Domain walls with
thickness much wider than $l_s$, is needed in the scenarios of
Sec.~\ref{subsubsec:CR_alternative}.

\subsection{$\mu_\mathrm{brane} \sim \mathbf{T}_3$?\label{subapp:brane_mass_density}}

The relationship $\mu_\mathrm{brane} \sim \mathbf{T}_3$ is used in
Sec.~\ref{subsec:fundamental_parameters}, to consider whether the
domain walls are relativistic or non-relativistic. By the Newton's
second law, the transverse wave equation of a stretched
(one-dimensional) string gives $v = \sqrt{T/\mu}$, where $T$ is the
tension and $\mu$ is the mass per unit length. Thus one possesses
the relation $\mu \sim T$ if the wave is relativistic. $\mu \sim T$
is also a general assumption in the study of cosmic
strings/superstrings.

\subsection{$T_\mathrm{dw} \sim \mathbf{T}_3^{3/4}$ or $T_\mathrm{dw} \sim \mathbf{T}_3 \mathbf{l}_p$?
    \label{subapp:holographic_tension}}

$T_\mathrm{dw} \sim \mathbf{T}_3^{3/4}$ or $T_\mathrm{dw} \sim
\mathbf{T}_3 \mathbf{l}_p$ can be obtained by dimensional analysis,
where $\mathbf{T}_3$ is the tension of the $(3+1)$-dimensional
branes perpendicular to the boundary of $\mathrm{AdS}_5$, and
$T_\mathrm{dw}$ is the holographic brane tension. It was used in the
estimations of Sec.~\ref{subsec:domain_wall_consistent}, in which
the domain walls are too heavy to be visible in our universe.
However, these two seem not the only answers.

On the one hand, similar to the RS I model, or the case of
Appendix~\ref{subapp:ls}, the right answer of $T_\mathrm{dw}$ should
depend on in what AdS radius the holographic domain wall visualizing
itself. The domain walls should be really light if the radius is
deeply inside the throat. On the other hand, as
Eq.(\ref{eqn:T3_tension}) gives $\mathbf{T}_3 \propto l_s^{-4}$,
$\mathbf{T}_3$ itself is much smaller in the (field) IR regions,
because the string length $l_s$ is much larger there. Hence,
$T_\mathrm{dw}$ is suppressed in that region, even if $T_\mathrm{dw}
\sim \mathbf{T}_3^{3/4}$ or $T_\mathrm{dw} \sim \mathbf{T}_3
\mathbf{l}_p$ is still valid. As discussed in
Sec.~\ref{subsec:domain_wall_consistent}, a smaller $T_\mathrm{dw}$
is needed, if we want another vacua to be visible, and the
contribution of the domain walls to the energy density of the
universe is consistent with the critical density.

\bibliographystyle{bib_style}
\bibliography{../0709_astro_particles,../0810_AdS-CFT,../0907_parameter,../0908_compactification}


\end{document}